# Electronegative metal dopants improve switching consistency in Al$_2$O$_3$ resistive switching devices

Zheng Jie Tan[1]*, Vrindaa Somjit[1]*, Cigdem Toparli[2], Bilge Yildiz[1,2,a)], and Nicholas Fang[3,b)]

[1]*Department of Materials Science and Engineering, Massachusetts Institute of Technology, Cambridge, Massachusetts 02139, USA.*

[2]*Department of Nuclear Science and Engineering, Massachusetts Institute of Technology, Cambridge, Massachusetts 02139, USA.*

[3]*Department of Mechanical Engineering, Massachusetts Institute of Technology, Cambridge, Massachusetts 02139, USA.*

* Equally contributing authors

Co-corresponding authors: a) byildiz@mit.edu, b) nicfang@mit.edu

**Abstract**

Resistive random access memories are promising for non-volatile memory and brain-inspired computing applications. High variability and low yield of these devices are key drawbacks hindering reliable training of physical neural networks. In this study, we show that doping an oxide electrolyte, Al$_2$O$_3$, with electronegative metals makes resistive switching significantly more reproducible, surpassing the reproducibility requirements for obtaining reliable hardware neuromorphic circuits. The underlying mechanism is the ease of creating oxygen vacancies in the vicinity of electronegative dopants, due to the capture of the associated electrons by dopant mid-gap states, and the weakening of Al-O bonds. These oxygen vacancies and vacancy clusters also bind significantly to the dopant, thereby serving as preferential sites and building blocks in the formation of conducting paths. We validate this theory experimentally by implanting multiple dopants over a range of electronegativities, and find superior repeatability and yield with highly electronegative metals, Au, Pt and Pd. These devices also exhibit a gradual SET transition, enabling multibit switching that is desirable for analog computing.

**Introduction**

Compact and energy efficient solid state resistive switching devices are actively being investigated as fundamental units for use as high-density non-volatile memories, and for enabling energy-efficient analog computing via physical neural networks[1,2]. These devices are capable of both data storage and computation, while being scalable to the nanometer regime[3,4]. This gives them great promise to circumvent the time and energy challenge of data movement[1] that plague current computing systems based on von-Neumann architecture with separate memory and computing units. One class of such devices is the resistive switching random access memory (RRAM)[5,6], which consists of a metal-insulator-metal (MIM) stack. The insulating solid electrolyte layer primarily made of chalcogenides or metal oxides becomes the switching medium. The reversible migration and redistribution of metals such as Ag or Cu[7], or of defects such as oxygen vacancies ($V_O$) through the electrolyte under the application of a voltage forms localized, tunable conductive regions that are responsible for switching[2,8]. The electrical modulation of the electronic conductance in an analog way in such resistive switching units is fundamental to brain-inspired analog computing[9].

Two key barriers preventing the widespread use of RRAMs are their high switching variability[3,4,10-13] and poor device yield[8,13-15]. These arise due to the inherent stochastic nature of the individual switching events. Variation in the location and the local chemistry and structure of such filaments leads to cycle-to-cycle and device-to-device variations in switching voltages and resistances, causing inconsistent switching[10,11,16]. In addition, pristine devices typically require an initial electroforming step, in which a voltage much higher than the set voltage is applied to form the first conductive path in the insulating electrolyte[17]. Such large forming biases can sometimes deform and destroy the devices[8,15], resulting in poor device yield. Poor switching repeatability and poor device yield adversely affect device stability, increase peripheral circuit complexity, and importantly, reduce computational accuracy of hardware-implemented neural networks, as highlighted by Gokmen et al.[18] and Li et al.[19]

Correspondingly, multiple strategies have been attempted to improve the switching consistency and yield of RRAM devices. For example, multilayer structures (such as $AlO_x/HfO_x$[20,21], $TiO_x/Al_2O_3$[22] and $HfO_x/TiO_x/HfO_x/TiO_x$[23]) are thought to enhance switching consistency by confining the filament formation and rupture pathways within very thin oxide layers[20-23].

Interdiffusion among the oxide layers and potential short-circuits across thin films of 1 to 2 nm thickness are limiting factors[22,24] to this approach. Nanocrystals[25,26] in the electrolyte (such as Ru[25] and Ag[26] nanocrystals in $Al_2O_3$) enhance the local electric field, and preferentially accelerate $V_O$ migration and cation dissolution, thereby reducing the randomness in filament formation, but to fabricate such nanostructures is not trivial or inexpensive. Other experiments suggest that introduction of metal dopants[14,27-32] (such as Ge[28]- and Al[27]-doped $HfO_2$) into oxide electrolytes improves switching consistency. First principles calculations reason that reduced $V_O$ formation energies near the dopant[28,29,31-35] increases switching uniformity by localizing the current path. No study, however, has systematically investigated the effect of the dopant properties on the performance of oxide devices using both experimental and simulation methods. Moreover, many of these devices still need high electroforming voltages[20,22,26,29]. Significant variability is still observed[20-23,25,27-29] and current compliances[22,25,27,29,30] are still used.

In this study, we focus on the link between $V_O$ formation and the electronegativity of dopants in an insulating oxide. Dopants such as Au, Pt, Pd, Rh on oxides are known to catalyze several important reactions, such as CO oxidation, water-gas shift, and NO reduction[36]. These dopants weaken metal-oxygen bonds in the host oxide lattice[36] while assisting these surface reactions. Following this, we hypothesize that dopants with high electronegativity can give rise to higher switching consistency in $Al_2O_3$ by acting as preferential sites for $V_O$ formation. Highly electronegative dopants reduce the formation energy of oxygen vacancies, because they weaken the metal-oxygen bonds, and create in-gap states and capture the electrons resulting from neutral oxygen removal (a process that is energetically impossible in undoped $Al_2O_3$).

In this study, we combine the above advantages of multilayer thin films and metal dopants to develop a device with superior switching consistency and high yield, that requires no external control circuitry, and is electroforming-free. Our original device consists of alternating layers of $Al_2O_3$ and WN deposited on Si, with the highly electronegative Au as the top electrode material. $Al_2O_3$ and WN have lower interdiffusion tendency[37], which is an important improvement from previous device configurations[22]. Au atoms were implanted into the $Al_2O_3$ electrolyte as dopants during focused ion beam (FIB) milling to define the device area. Density functional theory (DFT) calculations revealed a significant lowering of $V_O$ formation energy in the vicinity of Au, due to the changes in the electronic structure brought about by Au's high electronegativity. This process guides the formation of conducting paths, resulting in higher switching consistency. Multilayer devices doped in this way with Au had superior cycle-to-cycle and device-to-device switching consistency than the undoped devices, consistent with the prediction of stable conducting paths obtained by the DFT model. Based on this understanding, we predict and validate other highly electronegative dopants such as Pt and Pd to increase uniformity of resistive switching among multiple devices and cycles, as compared to active transition metals as Cu, Ti and Al. Furthermore, our device exhibits a gradual SET transition, which, coupled with its high uniformity, makes it a favorable candidate for use in multibit switching applications.

**Results**

**Effect of Au doping on device switching consistency**

We tested our hypothesis about the effect of dopant electronegativity on switching consistency first on Au-doped $Al_2O_3$. Au is one of the most electronegative metals in the periodic table[38]. More than 100 multilayer RRAM devices were fabricated and tested. These RRAM devices were made of alternating layers of $Al_2O_3$ separated by conductive WN layers, with Au as the top electrode and WN as the bottom electrode. The device cross-section imaged using a transmission electron microscope (TEM) is shown in Fig. 1a. The device schematic, and the effect of Au doping on switching consistency is shown in Fig. 1b-f. From the I-V plot of over 300 switching cycles in Fig. 1b, it is seen that depositing the Au top electrode after FIB milling results in inconsistent switching. In contrast, in Fig. 1c, it is clear that FIB milling to define the device area after depositing the Au top electrode enhances the switching consistency dramatically. Estimations from Stopping and Range of Ions in Matter (SRIM) simulations indicate that the high-energy FIB milling process results in the implantation of Au atoms from the top electrode into the $Al_2O_3$ electrolyte (Supplementary Fig. 1-3). This being the only difference between the two schemes, confirms that Au implantation leads to the difference in switching consistency. Alternative doping procedures, such as co-sputtering of Au during reactive sputtering of $Al_2O_3$ layers were also explored, with similar results (Supplementary Fig. 4). Fig. 1d and 1e show the cumulative distribution function (CDF) plots of the obtained high- and low-resistance states (HRS and LRS) at the device-level and at the cycle-level, respectively. To evaluate the variability in LRS and HRS, we use a new and more reliable measure of switching consistency, the logarithmic coefficient of variation $(C_{lv})$[39], defined as the difference between the 10$^{th}$ and 90$^{th}$ percentile of the logarithm of resistance values.

Compared to $C_{lv}$ ~ 1.5 for other Al$_2$O$_3$-based RRAMs reported in literature[40,41], our devices have significantly lower $C_{lv}$ of 0.1 and 0.34 for the LRS and HRS respectively, as plotted in Fig. 1d. This slight switching variability arises mainly from fabrication-related variation, which can in fact be further reduced by enforcing stringent manufacturing procedures. From Fig. 1e, it is clear that cycle-to-cycle variation has an even smaller spread of about 0.04 and 0.05 for the LRS and HRS states, respectively. Thus, the intrinsic switching variability of each device is very low, indicating substantial reduction in the stochasticity of filament formation and rupture. Such low variability is valuable for facilitating multibit switching schemes[13], and meets and surpasses the reproducibility requirement (with a $C_{lv}$ of about 0.32) needed to implement accurate hardware neural networks[18]. Additionally, these multilayer, Au-implanted devices exhibit a perfect yield. All the devices are in LRS upon fabrication, precluding the need for electroforming (shown in Supplementary Fig. 15). This is advantageous because electroforming-free devices eliminate the need of peripheral circuitry in RRAM arrays, and high yield leads to improved accuracy of neural networks[19]. Moreover, the devices exhibit resistive switching with a gradual SET transition. This gradual SET transition is key to multibit switching and analog processing, as explained in forthcoming sections.

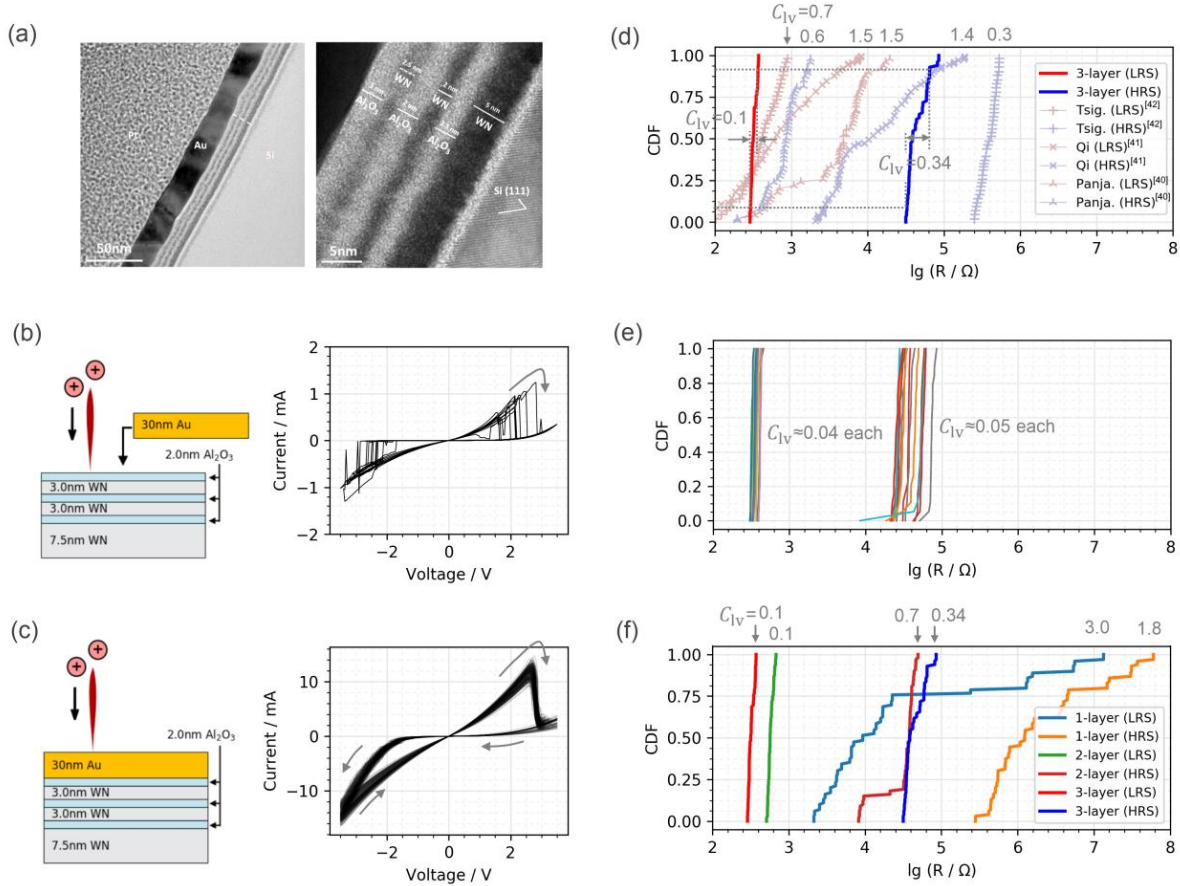

**Figure 1. Electronegative Au dopants and multilayering of the oxide films are two key factors to enable consistent switching. (a)** Device cross-section imaged using transmission electron microscopy (TEM). For further elaboration on device design and TEM imaging, see Supplementary Figs. 12 and 13. **(b)** Multilayer $Al_2O_3$ cannot switch consistently when FIB milling is performed to define the device area before Au deposition. **(c)** Multilayer $Al_2O_3$ switches consistently when FIB milling is performed to define the device area after Au deposition. These devices did not require electroforming. The I-V plot shows 300 superimposed switching cycles (15 devices x 20 consecutive cycles in each device). **(d)** The consistency improvement using the Au doping strategy can be seen from the short span of resistances in the cumulative distribution functions (CDFs), observed over the 300 switching cycles, covering only 0.10 and 0.34 decades for the LRS and HRS, respectively. Typical spans of other $Al_2O_3$[40,41] and $TiO_2$[42] devices from literature are shown for comparison. **(e)** The

**device-to-device CDF in (d) can be displayed separately for the 15 devices to show cycle-to-cycle variations smaller than 0.05, indicating that most of the variation seen in (d) comes from device-to-device differences. (f) CDF plots show the narrowing of the spread in the LRS and HRS resistances from a few orders of magnitude down to 0.10 and 0.34, respectively, with increasing the number of oxide layers from one to three.**

Switching from high resistance to low resistance state in our device is likely due to the formation of a network of filaments made of oxygen vacancies ($V_O$). The dependence of switching current on the FIB-processed device perimeter indicates areal switching, as shown in Supplementary Figure 3. This indicates the presence of a network of conducting zones near the periphery of the device. This conducting zone is potentially in the form of a network of conducting filaments; or in the limit of a dense network, it is made of a host chemistry with higher conductance. It is unlikely that conducting zone is made of Au filaments via dissolution of the Au top electrode. Au is resistant to oxidation - and is in fact commonly used as the inert electrode in RRAM devices, as opposed to metals like Ag and Cu, which are typical active electrode candidates[43,44]. Additionally, metallic filaments typically result in a high ON-OFF ratio in the $10^4$-$10^8$ range[43], contrary to the ON-OFF ratio of 100 observed in our device. Furthermore, $Al_2O_3$ is a common electrolyte used in RRAM devices, which switches via $V_O$ filament formation as reported in various studies[20,26,45,46]. Thus, we believe that the role of Au is through its effect on $V_O$ formation in $Al_2O_3$ rather than through Au metal filament formation.

High switching consistency is observed in only multilayer devices doped with Au. The CDF plot for one-, two- and three-layer devices in Fig. 1f shows that the switching consistency improves as the number of layers is increased from one to three, while keeping the total thickness of $Al_2O_3$ constant. This is likely because thinner oxide layers require shorter conducting paths, thereby lowering stochasticity in the formation and disruption of conductive channels of oxygen vacancies bridged by Au dopants, giving rise to high uniformity as observed in previous studies on multilayer oxides[20-23]. Thus, having a multilayer and Au-doped electrolyte together form a necessary and sufficient condition to achieve high switching consistency in our system.

**Effect of Au doping on switching consistency: model based on first principles calculations**

We have carried out density functional theory (DFT) calculations to identify the effect of Au doping on oxygen vacancy ($V_O$) formation. Irradiation processes such as FIB milling result in the creation and distribution of defects such as vacancies and interstitials. Here, we focus on the effect of Au at the interstitial site in $Al_2O_3$. The effect of Au at the substitutional site on Al is considered in Supplementary Table 1 and follows similar trends. The concentration of dopants in our simulations is ~4%. Table 1 shows the formation energy of $V_O$ in undoped $Al_2O_3$ and at the nearest neighbor site of the interstitial Au dopant in $Al_2O_3$. The formation energy of $V_O$ next to the Au dopant is drastically lowered, by over 6 eV. This indicates that $V_O$ preferentially forms at the vicinity of the Au interstitial dopant. Since the Au atoms are pinned and stationary, the locations at which $V_O$ is formed are also fixed. This minimizes the randomness in $V_O$ formation, thus creating defined local regions that are easily reduced, which then connect to form conducting paths.

| Defect | Formation energy (eV) | Defect cluster type | Formation energy (eV) | Binding energy (eV) |
|---|---|---|---|---|
| $V_O$ | 7.19 | $V_O$ cluster | 7.01 | 0.18 |
| $V_O$ NN to Au | 0.64 | $V_O$ cluster with Au | 4.32 | 2.87 |
| $V_O$ NN to Pt | 0.47 | $V_O$ cluster with Pt | 4.73 | 2.48 |
| $V_O$ NN to Pd | 1.91 | $V_O$ cluster with Pd | 4.73 | 2.48 |
| $V_O$ NN to Cu | 3.61 | $V_O$ cluster with Cu | 5.29 | 1.90 |
| $V_O$ NN to Ti | 4.38 | $V_O$ cluster with Ti | 5.36 | 1.83 |
| $V_O$ NN to Al | 3.63 | $V_O$ cluster with Al | 5.40 | 1.79 |

**Table 1. (left) Oxygen vacancy ($V_O$) formation energies in undoped $Al_2O_3$ and at the nearest neighbor (NN) site of the interstitial dopants (Au, Pt, Pd, Cu, Ti, and excess Al) in doped $Al_2O_3$; (right) Oxygen vacancy ($V_O$) cluster formation energies and binding energies (per $V_O$) in undoped and doped $Al_2O_3$.**

In addition to the formation ease of $V_O$ point defects, it is important to investigate the effect of the Au dopant on $V_O$ cluster formation. $V_O$ clusters act as building blocks for the formation of conducting paths or networks of conducting channels via which resistive switching occurs. Table 1 shows the calculated $V_O$ cluster formation energies in undoped and Au-doped $Al_2O_3$. Cluster formation energy (per $V_O$) represents the ease of forming a $V_O$ cluster in the presence of Au. It is clear that introduction of Au dopant into the cluster markedly lowers cluster formation energy.

The reduction in the formation energy of $V_O$ next to the Au dopant can be rationalized by investigating the density of states (DOS) and electron redistribution of the doped $Al_2O_3$ system. Oxygen vacancy formation in undoped $Al_2O_3$ is an energetically costly process because the electrons that are left behind upon removing an oxygen atom cannot occupy the high energy, empty cation states and consequently, localize at the oxygen vacancy site. In contrast, in the DOS plots of the Au-doped $Al_2O_3$ shown in Fig. 2a i and ii, it is clear that the Au interstitial introduces additional states at the valence band maximum (VBM) of $Al_2O_3$, along with mid-gap states. The low-lying mid-gap states trap the electrons left behind upon the removal of an oxygen atom, completing the Au 6s orbital electron configuration ($[Xe]4f^{14}5d^{10}6s^1$ → $[Xe]4f^{14}5d^{10}6s^2$). The ability to uptake these electrons to the low-energy states decreases the $V_O$ formation energy. This capture mechanism of electrons from $V_O$ by the Au atom can be seen in the DOS plot in Fig. 2a ii, where the Au mid-gap states shift lower in energy. The corresponding partial charge density plot of the mid-gap states is shown in Fig. 2b, where an electron cloud around Au is clearly seen.

Additionally, Au is a noble metal with a high electronegativity of 2.3[38]. This leads to electron redistribution from Al to Au in $Al_2O_3$, facilitated by Au electronic states near the VBM of $Al_2O_3$ as noted above and shown in Fig. 2a i and ii. The calculated Bader charge[47,48] on Au, nearest-neighbor Al and nearest-neighbor O in Au-doped $Al_2O_3$ is -0.4 e, +2.41 e, and -1.55 e, respectively. The magnitude of charge on nearest-neighbor O is lower than in the undoped case, where O has a charge of -1.65 e. The presence of Au dopant thus leads to electron transfer from Al to the Au atom instead of to O. Charge transfer from Al to Au weakens the nearest-neighbor Al-O bonds in the $Al_2O_3$ lattice, resulting in the lowering of the $V_O$ formation energy. Electron transfer from $Al_2O_3$ to Au dopant has been observed experimentally in prior work[49,50], for example in Au-$Al_2O_3$ nanocomposites[50] as well as upon adsorption of Au monomers on $Al_2O_3$/NiAl[49].

The initial structure of the Au-doped $V_O$ cluster with Au at the interstitial site in $Al_2O_3$ is shown in Fig. 2d, with the positions of $V_O$ marked with black circles. The oxygen vacancy cluster introduces multiple discrete mid-gap states, shown in Fig. 2b i and ii. Introducing Au gives rise to additional mid-gap states, particularly near the top of the valence band and the bottom of the conduction band. As seen in Fig. 2b, the dominant mid-gap states are from $V_O$, and these states can provide a path for electrons to tunnel through the oxide barrier from the cathode to the anode[11]. It is expected that, as the concentration of oxygen vacancies increases under applied field, the number of localized states from $V_O$ in the band gap will increase, ultimately closing the band gap, giving rise to metallic conduction. In fact, this is seen in the density of states plots of the Au-doped vacancy filament path model in Supplementary Fig. 6 and 7 (additional details regarding the filament model can be found in Supplementary Tables 2-4). The partial charge density of all the defect states within the band gap of the relaxed, Au-doped system with $V_O$ cluster is shown in Fig. 2e, revealing a localized, conductive cluster arising from the states introduced by oxygen vacancies.

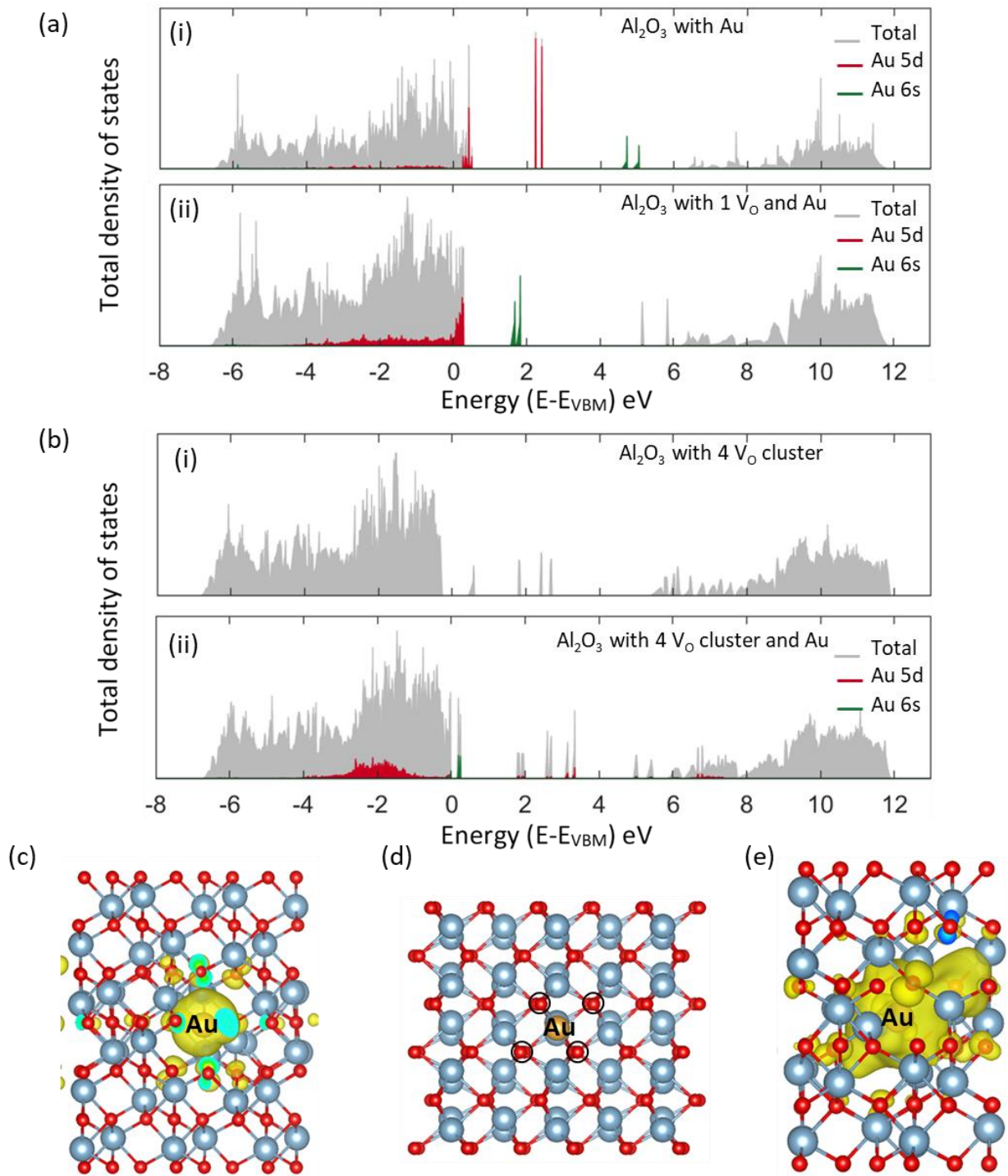

**Figure 2. (a)** Total density of states (DOS) of Au-doped $Al_2O_3$ (i) and of Au-doped $Al_2O_3$ with one $V_O$ at nearest neighbor site to Au (ii). **(b)** Total DOS of undoped $Al_2O_3$ with a 4-$V_O$ cluster (i) and Au-doped $Al_2O_3$ with a 4-$V_O$ cluster (ii). In all DOS plots, valance band maximum is at 0 eV. **(c)** Band decomposed charge density profile of Au s-orbital in relaxed Au-doped

Al$_2$O$_3$ with a single V$_O$ (Isosurface: 0.003 eV/ Å$^3$), showing charge transfer to Au. (d) Initial structure of the V$_O$ cluster in Au-doped Al$_2$O$_3$ (positions of the nearest neighbor V$_O$'s are marked by black circles). (e) Band decomposed charge density profile of electronic states within the bandgap for the relaxed Au-doped Al$_2$O$_3$ with a 4-V$_O$ cluster around Au (Isosurface: 0.01 eV/Å$^3$).

Next we assess the V$_O$ cluster binding energies which are reflective of the preferential position of vacancy or vacancy cluster formation. Table 1 tabulates the V$_O$ cluster binding energies in undoped and Au-doped Al$_2$O$_3$. Cluster binding energy (per V$_O$), calculated as $E_b(nV_O - D) = \left(nE_f^{Vo} + E_f^D - E_f^{nVo-D}\right)/n$, represents the energy required to dissociate the cluster into isolated V$_O$ and Au. The binding energy of the V$_O$ cluster in the undoped case as calculated in this study matches well with previous theoretical studies on V$_O$ chains[46] and V$_O$ pairs[51]. It is clear that the presence of the Au dopant increases cluster binding energy (per V$_O$). The positive binding energy indicates that the cluster is cohesive. Thus, introduction of Au not only makes V$_O$ cluster formation more energetically favorable, but also enhances cluster cohesion. Such short-range cohesive clusters can then act as building blocks for the formation of conducting paths across the entire oxide layer. When Au is present, formation of vacancies and vacancy clusters preferentially occur near the Au atoms, thereby, reducing stochasticity in the formation of conducting paths and increasing the cohesion and stability of these paths.

Given the favorable formation and binding energies of V$_O$ and V$_O$ clusters near the Au dopant, a reasonable question arises about whether this binding leads to a reduction in mobility of V$_O$. As shown in Supplementary Figs. 8-10 and Supplementary Tables 5 and 6, we have calculated V$_O$ migration barriers in the Au-doped system. In all the paths studied, we find that the migration barriers are lower by ~0.5 eV compared to that in undoped Al$_2$O$_3$. Thus, V$_O$ is more mobile in the Au-doped system as compared to the undoped system. This is due to charge transfer from V$_O$ to Au, reducing the trapped electron density in V$_O$, and thereby making the migration of V$_O$ easier than in the undoped Al$_2$O$_3$. These results reveal favorable implications for switching speed and energetics via V$_O$ formation and migration in Au doped Al$_2$O$_3$.

**Prediction of other dopants and their device tests**

Given the above proposed connection between dopant electronegativity and switching consistency as explained in the previous section, we have assessed the effect of more and less electronegative dopants on the $V_O$ point defect and $V_O$ cluster formation energies, and on device switching repeatability. A range of dopants across the electronegativity scale[38] was studied, namely, Pt (2.1), Pd (2.0), Cu (1.8), Ti (1.6) and Al (1.5) interstitials. This expands the device design space as well as further strengthens the link between dopant electronegativity and device variability.

As seen in Table 1, while the formation energy of nearest-neighbor $V_O$ is generally lowered regardless of the interstitial dopant, for the highly electronegative Pt and Pd dopants, the formation energy is very significantly lowered. Thus, similar to Au, electronegative dopants like Pt and Pd also reduce the $V_O$ point defect and cluster formation energy considerably, and have higher cluster binding energy. The relaxed structures of $V_O$ at the nearest neighbor site of these interstitial dopants can be found in Supplementary Fig. 19, and the total DOS of the doped $V_O$ clusters can be seen in Supplementary Fig. 20. It is worthwhile to mention here that we have also studied the resulting change in the $V_O$ formation energy when these dopants were placed as substitutional at the Al site, and we have found similar trends as the interstitial dopants (Supplementary Table 1).

Bader charge analysis[47,48] revealed that, similar to the case with Au, charge transfer takes place from Al to Pt and Pd, but not to Cu, Ti or Al interstitial (Supplementary Table 7). Investigating the electronic DOS, low-lying states near the VBM and mid-gap states are observed in the Pt- and Pd-doped cases as well, but not in the Al-, Cu- and Ti-doped cases (see Supplementary Fig. 11 for local DOS plots for all cases). The states near the VBM facilitate electron redistribution around the electronegative Pt and Pd dopants, easing Al-O bond breakage. The mid-gap states trap the electrons left behind upon removal of an oxygen atom, thus, lowering $V_O$ formation energy significantly.

It is noteworthy to point out here that while the difference between the electronegative and non-electronegative elements is clear, the relative trend between the electronegative dopants can also be explained. It is seen that $V_O$ defect formation energy next to Au and Pt is lower than Pd, by over ~1 eV. This can be attributed to the relativistic contraction of the s- and p-orbitals of Au and Pt, due to their significantly higher mass compared to Pd. As explained by Pyykko and Desclaux[52], this contraction leads to the 6s state of Pt and Au lying deeper in the atom (as compared to that in Pd[53]), resulting in significant energy gains upon filling it.

We have validated these predictions on the role of electronegativity of the dopant on creating preferential zones of higher conductivity, by performing switching experiments on these compositions. Multilayer devices with the same geometry as shown in Fig. 1, were fabricated with Pt, Pd, Cu, Ti and Al top electrode layers, and FIB milled after top electrode deposition to define the device area. The I-V curves and corresponding CDF plots are shown in Fig. 3. In line with our computational predictions, devices doped with noble metals with high electronegativities such as Pt and Pd exhibit markedly consistent switching behaviors, whereas the more reactive metals with lower electronegativities are seen to have erratic switching cycles. $C_{lv}$ of the devices doped with the more electronegative metals is almost two orders of magnitude lower than those doped with Cu, Ti and Al.

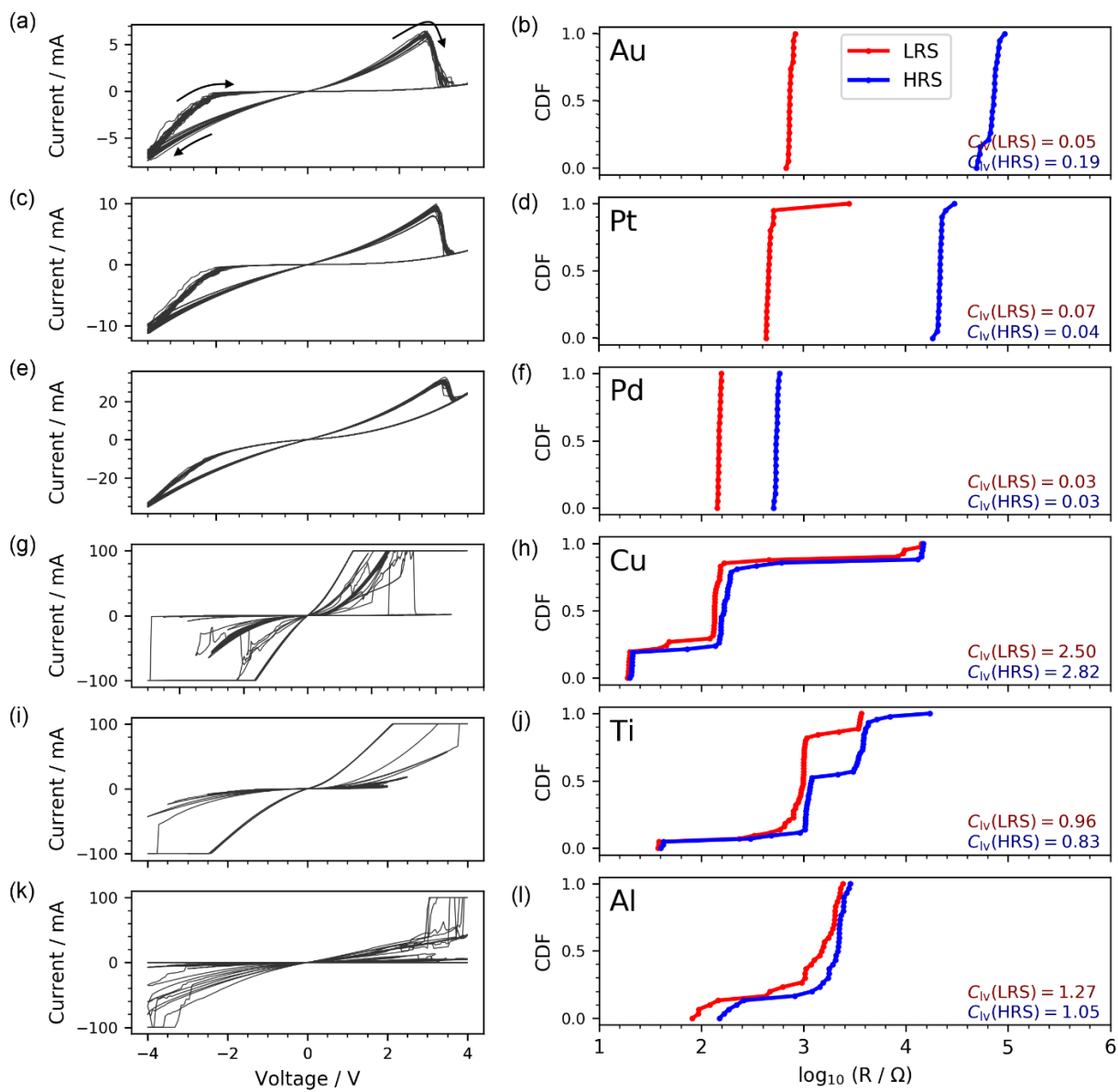

**Figure 3.** Dopant metals from a range of electronegativities- Au, Pt, Pd, Cu, Ti, Al (corresponding to (a), (c), (e), (g), (i), (k))- were tested to observe their effects on switching consistency of $Al_2O_3$. The cumulative distribution function for the LRS and HRS resistances are plotted in (b), (d), (f), (h), (j), (l). Dopants with higher electronegativities (Au, Pt, Pd) have CDF plots with narrower widths, indicating that $Al_2O_3$ layers doped with these metals have more consistent switching, consistent with predictions of easier Vo and Vo cluster formation shown in Table 1 and Figure 2. Dopants with lower electronegativities (Cu, Ti, Al) exhibit erratic switching, and with poor ON-OFF ratios.

## Multibit switching

This high switching consistency, demonstrated by Au, Pd and Pt dopants in $Al_2O_3$ is beneficial for achieving multibit switching. From the I-V plots in Fig. 4 (and also in Fig. 3a,c,e), it can be observed that these doped devices exhibit a gradual SET transition. The gradual SET transition allows the modulation of resistance states in a continuum manner, a key requirement for analog computing. The choice of a different terminating cycle voltage in the voltage-sweep measurements leads to different final resistance states, with distinct I-V traces, as seen in Fig. 4a. A more negative terminating cycle voltage puts the device in a more conductive LRS state (plotted in Fig. 4b and 4c), and subsequently also has a RESET transition that occurs at a larger voltage. The gradual increase in conductivity during SET is likely via the increase in the volume of conductive pathways[11]. The larger the change in the resistance state upon set, the larger the positive reset voltage that was needed, as also seen in Fig. 4a. It is worthwhile to point out here that this multibit switching is demonstrated using a blind strategy, i.e., without any feedback control to read the resistance state and make adjustments. Additionally, no external control circuitry was used to enforce a SET current compliance. This simplifies the circuit design significantly, which will be useful in reducing the effective footprint of each cell for future multibit RRAM arrays. Additional details regarding this scheme are elaborated in Supplementary Figs. 17 and 18. This feature of multilevel resistance states, along with the superior switching consistency, makes these devices favorable candidates for multibit resistive switching. The multibit switching exhibited here could also extend the range of programming options for neuromorphic computing applications which currently relies on voltage or current pulses to update each device.

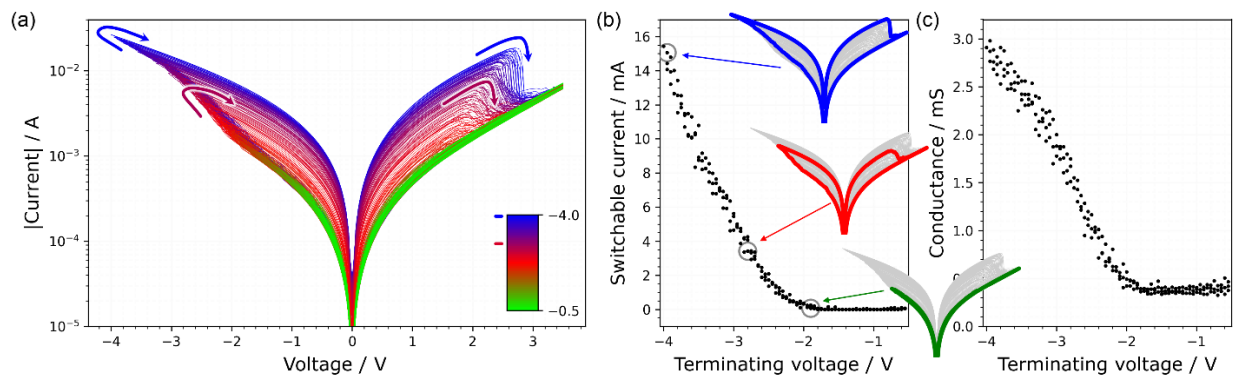

**Figure. 4.** The gradual SET process is convenient for demonstrating multibit switching in $Al_2O_3$. In (a), multiple switching cycles were performed with variable terminating voltages between -0.5 to -4V, and a fixed maximum voltage at 3.5V. Each switching cycle is color-coded according to that minimum terminating voltage used. For example, a voltage sweep from -4V to 3.5V is shown in blue, has the largest hysteresis that indicates the extent of switching, and the sharpest RESET onset at about 2.8V. On the other hand, a voltage sweep from -1V to 3.5V is shown in green and has barely any hysteresis or switching. Subplot (b) and (c) shows that the simple use of a chosen terminating voltage is able to put the device into a predictable state as characterized by the switchable current in (b) and the device conductance in (c). The switchable current is defined as the difference in the device current as set and the lowest current measured in the OFF (highest resistance) state. The green, red and blue overlays in (b) show the nature of one I-V sweep when the device is set by -2V, -2.8V and -3.9V as terminating voltages, respectively. No current compliance was needed to be programmed in the sourcemeter used for these measurements.

## Conclusion

In conclusion, this study identified that doping the insulating oxide electrolyte in the RRAM device with electronegative metal dopants can significantly improve the device switching consistency. Our computational analysis reveal that electronegative dopants act as preferential sites for the formation of $V_O$ point defects and clusters, and also increase the binding energy of the $V_O$ clusters in $Al_2O_3$. This is because the midgap states introduced by the electronegative metal dopants capture the electrons left behind upon removal of oxygen, and also weaken Al-O bonds, facilitating $V_O$ formation. These clusters then act as building blocks for the formation of networks of conductive and cohesive $V_O$ filaments. Thus, electronegative dopants reduce the number of possible filament pathways and thereby increase the uniformity of each device. Additionally, the mid-gap states that are introduced dominantly by the $V_O$ clusters provide a path for easy electron conduction. The devices doped with electronegative dopants, Au, Pt, Pd, have a cycle-to-cycle variation of just ~0.05 in log-scale for the HRS and LRS, and have a logarithmic coefficient of variation almost two orders of magnitude lower than those doped with active elements, Cu, Ti, Al. This high uniformity, coupled with the gradual SET transition of the device, was used to demonstrate multibit switching capability without any external circuitry. Thus, this work enables the development of a high yield, electroforming-free RRAM device with minimal footprint, superior switching consistency and multibit capability, while also providing mechanistic insights into strategies to improve device uniformity. This will benefit efforts in RRAM device design and integration into crossbar arrays for use in neuromorphic computing applications.

**Methods**

**A. Density functional theory calculations**

The energetics of the 2x2x1 perfect supercell (a=9.62 Å, c=13.13 Å) and of all the defects were calculated using density functional theory (DFT) using a plane-wave basis set, projector-augmented wave pseudopotentials[54] and the Perdew-Burke-Ernzerhof (PBE) parameterization of the generalized gradient approximation (GGA)[55] as the exchange-correlation functional, as implemented in the Vienna ab initio Simulation Package (VASP)[56] v.5.4.1. A kinetic energy cutoff of 520 eV and a gamma-centered 2x2x2 k-point mesh was used, resulting in a convergence accuracy of < 1meV/atom. All calculations were performed with a Gaussian smearing width of 0.05 eV and spin-polarized setting. Atomic positions were relaxed until the force on each atom was below 0.02 eV/Å.

The formation energy of a neutral oxygen vacancy (V$_O$) in bulk Al$_2$O$_3$ was calculated as

$$E_f = E_{DFT}^{V_O} - E_{DFT}^{perf} - \mu_O \tag{1}$$

where $E_f$ is the formation energy of V$_O$ in bulk Al$_2$O$_3$; $E_{DFT}^{V_O}$ is the DFT energy of the supercell with a V$_O$; $E_{DFT}^{perf}$ is the DFT energy of the perfect supercell with no defects, and $\mu_O$ is the chemical potential of oxygen in the system, calculated in the oxygen-rich limit as given in Equation 3.

The formation energy of V$_O$ nearest neighbor (NN) to a dopant, with the dopant occupying the octahedral interstitial site of Al$_2$O$_3$) was calculated as

$$E_f = E_{DFT}^{V_O-D} - E_{DFT}^{D} - \mu_O \tag{2}$$

where $E_f$ is the formation energy of V$_O$ at the NN site to the dopant; $E_{DFT}^{V_O-D}$ is the DFT energy of the supercell with the dopant and V$_O$ at its NN site; $E_{DFT}^{D}$ is the DFT energy of the supercell with the dopant at the interstitial site, and $\mu_O$ is the chemical potential of oxygen in the system, calculated in the oxygen-rich limit, i.e.,

$$\mu_O(T, P_{O_2}) = \frac{1}{2}\left[E_{O_2}^{DFT} + E_{over} + \mu_{O_2}^0(T, P^0) + kT\ln\left(\frac{P_{O_2}}{P^0}\right)\right] \tag{3}$$

where $E_{O_2}^{DFT}$ is the DFT energy of the O$_2$ molecule, $E_{over}$ is the correction for the O$_2$ overbinding error caused by GGA, taken as 1.36 eV as identified by Wang et al.[57] $\mu_{O_2}^0(T, P^o)$ is the difference in chemical potential of O$_2$ gas between $T = 0\ K$ and the temperature of interest, at a reference pressure of $P^o = 1\ atm$, as obtained from thermo-chemical tables; $P_{O_2}$ is the partial pressure of oxygen gas ($1\ atm$ in the O-rich limit).

The cluster formation energy (per $V_O$) in undoped $Al_2O_3$ was calculated as

$$E_f = \left(E_{DFT}^{nV_O} - E_{DFT}^{perf} - n\mu_O\right)/n \tag{4}$$

and in doped $Al_2O_3$, it was calculated as

$$E_f = \left(E_{DFT}^{nV_O-D} - E_{DFT}^{D} - n\mu_O\right)/n \tag{5}$$

where $E_f$ is the formation energy of $V_O$ cluster in the undoped and doped case respectively, $E_{DFT}^{nV_O}$ is the DFT energy of the supercell with only the $n$ $V_O$ ($n$ = 4) cluster, $E_{DFT}^{nV_O-D}$ is the DFT energy of the supercell with the dopant and $n$ $V_O$ ($n$ = 4) cluster, $E_{DFT}^{D}$ is the DFT energy of the supercell with only the dopant at the interstitial site, $E_{DFT}^{perf}$ is the DFT energy of the perfect supercell with no defects, and $\mu_O$ is the chemical potential of oxygen in the system, calculated in the oxygen-rich limit as in Equation 3.

The cluster binding energy (per $V_O$) in the undoped case was calculated as

$$E_b(nV_O) = \left(nE_f^{V_O} - E_f^{nV_O}\right)/n \tag{6}$$

and in the doped case, it was calculated as

$$E_b(nV_O - D) = \left(nE_f^{V_O} + E_f^{D} - E_f^{nV_O-D}\right)/n \tag{7}$$

where $E_f^{V_O}, E_f^{D}$ have been defined before, and $E_f^{nV_O}, E_f^{nV_O-D}$ are the formation energies of the $V_O$ cluster in the undoped and doped cases respectively.

**B. ALD deposition of WN/Al$_2$O$_3$ stack**

N-type degenerate Si wafers were purchased from University Wafer. Wafers were dipped in 1:50 HF:H$_2$O for 60s to remove native oxide and spin rinsed dried. Next, a wafer was loaded into an Oxford FlexAL ALD machine for plasma-enhanced deposition of alternate layers of WN and Al$_2$O$_3$ to give the stack Si/7.5nm WN/2.0nm Al$_2$O$_3$/3.0nm WN/2.0nm Al$_2$O$_3$/3.0nm WN/2.0nm $Al_2O_3$. The bottom electrode of the resistive switching device is the 7.5nm WN. The purpose of this WN is to build the stack starting from a well-defined layer to avoid wafer to wafer variations from an uncertified supply of Si wafers. WN is used instead of other commonly used metals due to CMOS requirements imposed on this shared ALD machine. For a 2-layer or 1-layer $Al_2O_3$ device, the thickness of each oxide layer will be increased to 3.0nm and 6.0nm respectively so that the combined oxide layer thickness remains constant. Deposition was done at 300°C. The deposition of WN is a $N_2/H_2$ plasma-enhanced reaction with bis(tert-butylimino)bis(dimethylamino)tungsten(VI) (BTBMW) precursor. The deposition of $Al_2O_3$ is a $O_2$ plasma-enhanced reaction with trimethylaluminum (TMA) precursor. Both recipes were supplied by the manufacturer. The thin film thickness were determined via X-ray reflectivity using a Rigaku SmartLab X-ray diffractometer, with both single films on wafers or composite films on wafers measured. The growth rate of WN and $Al_2O_3$ was deduced to be 0.5Å/cy and 1.0Å/cy respectively on the Oxford FlexAL ALD machine.

## C. Au deposition and FIB ion milling

Au deposition is typically performed on a Balzers tabletop sputterer at 130V and 40mA for 150s to give a film thickness of 30nm. There is no difference in device performance when Au is instead being deposited with a AJA International magnetron sputterer or an e-beam deposition machine. No additional metal adhesion layer is used for Au deposition. A 30kV Ga ion beam on a FEI Helios NanoLab 600i DualBeam FIB/SEM was used to mill away material to produce a square mesa where each side of this square is 50μm and the width of the milled border is 1μm. The SEM mode was used to image the chip to set-up for the FIB so there is no unintended FIB damage except as intended around the perimeter of the mesa. The milling was performed to a depth that exposes the Si substrate. This corresponds to an areal Ga ion dose of 80 to 120pC/$μm^2$. Resistive switching devices start in the LRS after ion beam milling without a need for electroforming.

## D. Electrical measurements

Probing of the mesa was done with a 25μm diameter gold wire tip to contact the topmost Au film of a typical device on a custom-built probe station. The gold wire is soft and is great for avoiding scratches to the top film. A standard tungsten probe from Signatone (probe tip no. SE-T) can also be used. No difference in device performance was observed regardless whether the probe is an Au wire or a tungsten probe. The stiffer tungsten probe was necessary if the top film was Cu, Al, Ti and not Pt, Pd or Au because the Au wire is unable to punch through the native oxide of these metals. A Keithley 2450 sourcemeter was used to source voltage and measure current.

# Supplementary Materials

**Electronegative metal dopants improve switching consistency in Al$_2$O$_3$ resistive switching devices**

Zheng Jie Tan[1]*, Vrindaa Somjit[1]*, Cigdem Toparli[2], Bilge Yildiz[1,2,a)], and Nicholas Fang[3,b)]

# I. Estimation of sideward implantation profile

An estimation of the sideward implantation profile and concentration of Au dopants in $Al_2O_3$ as a side effect of FIB milling can be obtained with the help of Stopping and Range of Ions in Matter (SRIM) simulations.

There are two steps to the implantation process. First, 30 keV Ga ions impinge on the Au top electrode and some Au atoms are sputtered off this surface. Next, these sputtered Au atoms take off with some direction and some energy and have a chance of being implanted sideways into $Al_2O_3$ if it is ejected in the right direction and with a sufficient high energy.

To model the final step, SRIM simulation for 30 keV Ga ions into an Au film was performed to obtain the statistics for sputtered Au atoms (see Supplementary Figure 1). The distribution of the ejected Au atoms can be fitted noting that this fitting should match the SRIM data where sputtered Au atoms are more plentiful outwards normal to the surface and tend to have low energies rather than high energies.

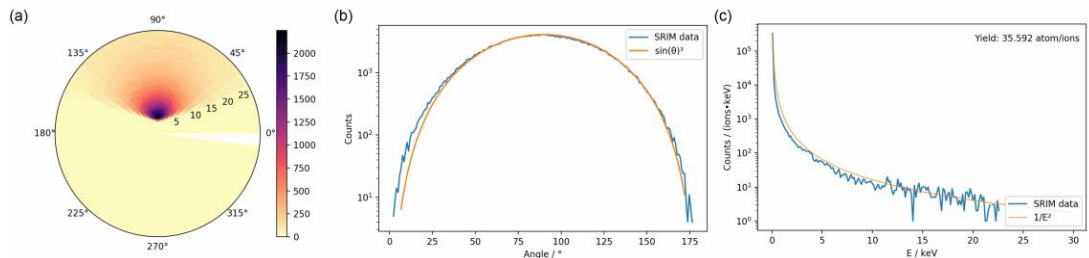

**Supplementary Figure 1.** (a) 2D distribution of sputtered Au atoms from Au film due to impinging 30 keV Ga ions. (b) The angular distribution of sputtered Au atoms can be plotted separately, with a good fit found to be proportional to $\sin^3(\theta)$. (c) The energy distribution of sputtered Au atoms can be plotted separately, with a good fit found to a $1/E^2$ profile.

For the second step, multiple sets of simulation was performed, each set for free Au atoms of a specific energy incident on $Al_2O_3$ at a specific angle, to find out the implantation profile of the Au atoms (see Supplementary Figure 2). The implantation depth can then be fitted with a Gaussian distribution which depends on the energy and incident angle as parameters.

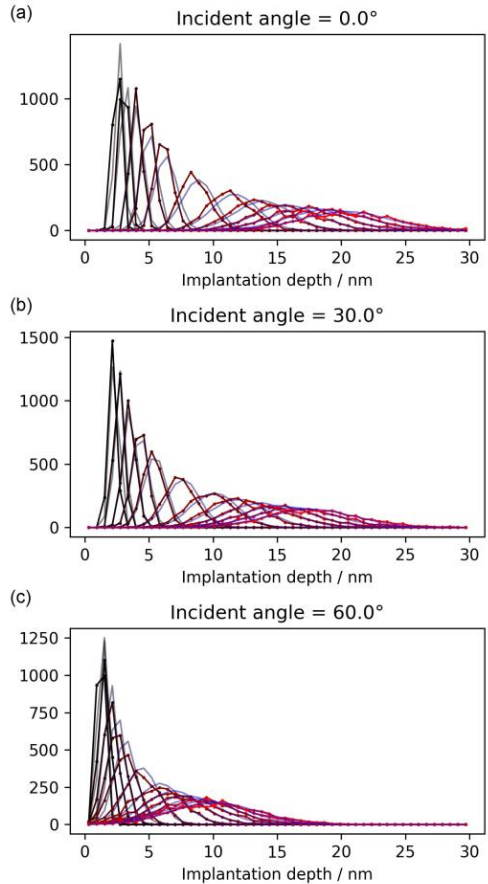

**Supplementary Figure 2.** SRIM simulations for various incident angles and ion energies of Au into $Al_2O_3$ were performed. (a), (b), (c) correspond to incident angles of 0, 30 and 60°. The lines vary from black to red, corresponding to a variation of energy from 0.1, 0.2, 0.5, 1, 2, 5, 10, 15, 20, 25, 30 to 35 keV. The pale blue curve behind each line corresponds to a model fit of the SRIM simulation data, showing that it is possible to obtain good model predictions of the SRIM outcome.

Finally, the two models can be combined for a rough estimate of the implantation depth of Au atoms into $Al_2O_3$ (see Supplementary Figure 3). However, this requires significant assumption of how both the microscopic and macroscopic morphology of the device changes during the process of the milling, so this means the implantation depth is good for a qualitative than a quantitative measure. Although the quantitative scale is uncertain, the overall trend is however expected to be reliable, so this means that some stray Au atoms are expected deep in $Al_2O_3$ with a decaying profile. The simulated implantation profile of Au atoms suggests that there is a large planar concentration of Au near the surface of $Al_2O_3$ after implantation.

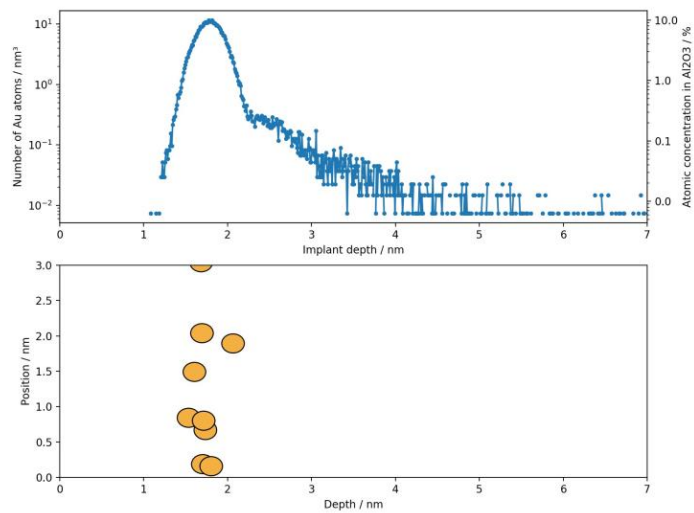

**Supplementary Figure 3.** The result from the implantation modelling can be plotted either numerically in (a) or graphically in (b) assuming a 2D cross-sectional view of the $Al_2O_3$ of one atom thick. The implantation assumes no interaction between implanted Au and results in a large number density of Au between 1.5 and 2.2 nm deep.

## II. Alternative fabrication techniques

A few other deposition requirements exist in addition to the need of multiple oxide layers for consistent resistive switching. The use of FIB in defining the device was further verified to be necessary. As a counter example, a chip deposited with the same repeating stack of WN and $Al_2O_3$ in the PEALD chamber was patterned with Au contact pads through a shadowmask using e-beam deposition. The devices were measured under a probe station and found to require large electroforming voltages of >10V, had wildly variable characteristics from device to device, and often failed by being permanently stuck in either the LRS or HRS.

We also identified process parameters which are relatively less important in contributing to the resistive switching behavior observed. The knowledge of which process parameters are of low importance is helpful in relaxing the conditions of device fabrication. The deposition method of the top Au layer is inconsequential as consistent switching is observed regardless whether the Au deposition is performed using e-beam deposition, DC or RF sputtering, or even using a tabletop Au sputterer. PEALD was used for the deposition of the repeated WN/$Al_2O_3$ stacks as the machine is an industrial grade tool in a strictly controlled process environment and can be reliably expected to produce repeatable depositions. We have recently migrated the recipe for deposition of the same stack with reactive sputtering in a RF/DC sputterer with some success. As a side note, thermal ALD of the WN/$Al_2O_3$ stack did not produce a switchable device. This turns out to be simply due to the bis(tert-butylimino)bis(dimethylamino)-tungsten precursor for WN not adhering on preceding $Al_2O_3$ layers but is potentially solvable by some novel modification of the thermal ALD recipe.

Lastly, we also have initial success with substituting the FIB milling treatment with co-sputtering of Au with $O_2$ reactive sputtering of Al to replicate the effects of Au implantation (see Supplementary Figure 4). The HRS and LRS of the Au co-doped device has a higher resistance than is typical of the usual FIB milled processed devices, but perhaps these resistances could be lowered by increasing the doping frequency or concentration of Au. This modification will eliminate the need to do expensive FIB milling processing or Au implantation into our device to activate the device for resistive switching. The use of co-doping to incorporate Au dopants into the $Al_2O_3$ layer will be a useful and independent parameter which can be independently used to adjust the HRS or LRS resistance as needed. In contrast, the magnitudes of the HRS and LRS resistances is more typically tuned by varying the thickness of the oxide in switching films but this would be an undesirable method because the switching behavior is largely dependent on the oxide thickness.

These results from alternative fabrication processes further strengthens our hypothesis that implanted Au atoms in $Al_2O_3$ is the key contributor for the highly consistent resistive switching behavior we observed.

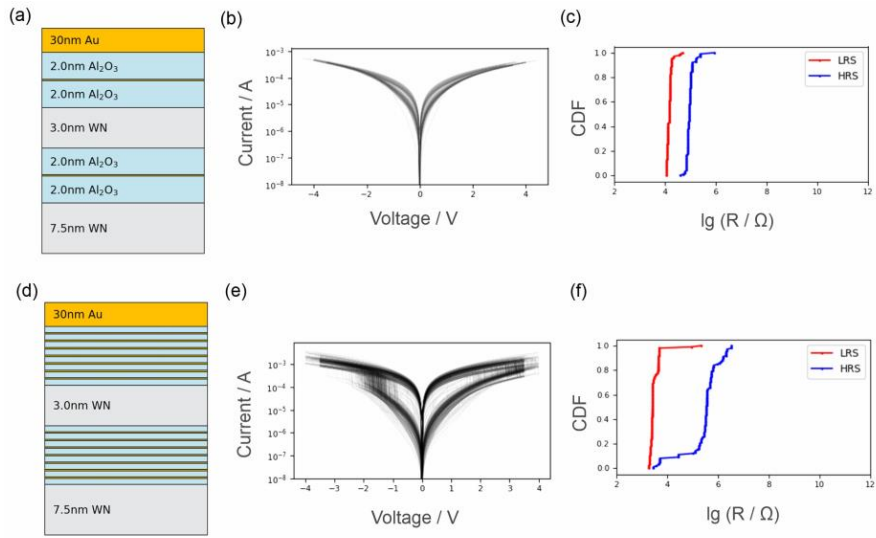

**Supplementary Figure 4.** (a) and (d) shows two examples of doping Au into 4 nm of reactively sputtered $Al_2O_3$ where there is one such Au dopant layer for each of the two $Al_2O_3$ layer in (a) while 7 such layers exist for (d). Each Au layer is the equivalent of 1 Å in deposition thickness but is expected to form Au islands upon deposition due to its thin thickness and the low wettability of Au on $Al_2O_3$. (b) & (c) show 157 switching cycles from 1 device while (e) & (f) show 870 cycles across 8 devices. The switching is fairly consistent but will require more optimization of the Au co- deposition parameters to more closely match the switching behaviors of the Au-implanted devices. Nonetheless, the improved switching behavior with Au co-doping is in support of our hypothesis.

**Supplementary methods for alternative reactive sputtered deposition**

As an alternative to ALD deposition, the WN and $Al_2O_3$ can also be deposited by reactive sputtering. WN was reactively sputtered with a 29 sccm Ar and 6 sccm $N_2$ mix at 3mT and DC 150W power on an AJA International Phase II J sputterer. $Al_2O_3$ was reactively sputtered with a 29 sccm Ar and 6 sccm $O_2$ mix, with plasma struck at 30mT and deposition at 3mT. The deposition rate of $Al_2O_3$ was 0.11 Å/s with a 300W RF power supply. If substituting FIB Au ion implantation, co-doping of Au can be done at this stage by flashing 2s to 5s of Au plasma at 20W DC for 3 to 7 times spaced evenly within each $Al_2O_3$ layer.

## III. Direct dependence on device properties with milling

Devices with e-beam deposited contact pads can be made to demonstrate consistent switching by FIB milling lines into them. Multiple devices with e-beam deposited contact pads of various sizes and milled lines of various lengths were created. The conductance of both of the LRS and HRS was found to vary linearly with respect to the length of the milled line while being independent on the device area (see Supplementary Figure 5). Also, the direct dependence on the resistive switching behaviour and the FIB milling and null dependence with device area indicates that this device can be scaled down significantly.

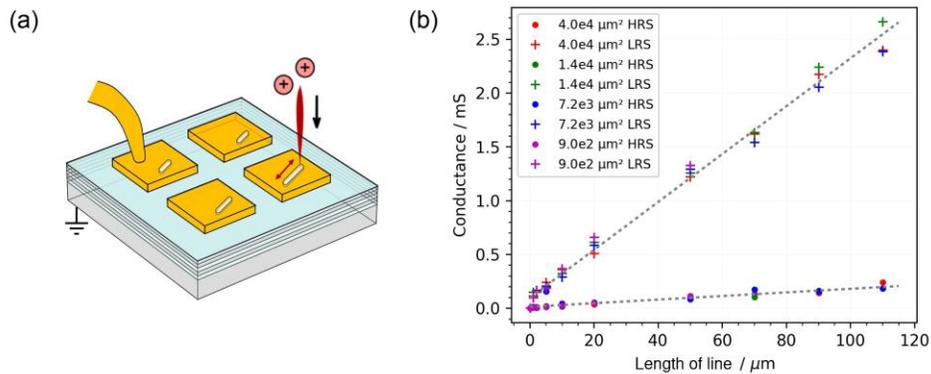

**Supplementary Figure 5. (a) A 30nm thick Au contact pad of variable size was deposited with e-beam deposition through a shadowmask onto the usual triply repeated WN/Al$_2$O$_3$ stack. A line with variable length was then milled into this Au contact pad with an FIB machine. (b) There is a linear relation of the obtained device conductances with the FIB milled lengths but no dependence on the contact pad area. This shows the device character is directly determined by the FIB milling and implies that the device area can be scaled down arbitrary.**

# IV. Effect of substitutional dopants on $V_O$ formation and binding energies in $Al_2O_3$

As shown in Supplementary Table 1, the effect of substitutional doping of the Al site on formation energy of a single neutral $V_O$ at the nearest neighbor site to the dopant follow a trend similar to the interstitial doped cases discussed in the main text. Noble metal dopants Au, Pd, Pt lower the $V_O$ formation energy more significantly than the more reactive Cu and Ti dopants.

| Defect | Formation energy $E_f$ (eV) |
|---|---|
| $V_O$ | 7.19 |
| $V_O$ NN to Au | 3.66 |
| $V_O$ NN to Pt | 4.14 |
| $V_O$ NN to Pd | 4.19 |
| $V_O$ NN to Cu | 4.34 |
| $V_O$ NN to Ti | 6.68 |

**Supplementary Table 1.** Oxygen vacancy ($V_O$) formation energies in undoped $Al_2O_3$ and at nearest neighbor (NN) site of substitutional dopants at the Al site in doped $Al_2O_3$

## V. V$_O$ filament model

In addition to V$_O$ clusters, V$_O$ filaments were investigated using DFT. We used the vacancy chain model as simulated previously in Al$_2$O$_3$[1], HfO$_2$[2] and TiO$_2$[3,4], i.e., consisting of 8 nearest neighbor V$_O$ along $[2\bar{1}\bar{1}0]$. This represents an ordered conductive filament. The dopant was inserted in an octahedral interstitial site at the center of the chain. Supplementary Table 2 shows the calculated V$_O$ chain formation energies and binding energies in undoped and doped Al$_2$O$_3$. The binding energy of the V$_O$ chain in the undoped case as calculated in this study matches well with previous theoretical studies[1,5]. It is apparent that the trend of the electronegative dopants Au, Pt and Pd lowering the formation energies (and increasing the binding energies) more than Cu, Ti and Al is still followed.

| Chain type | Formation energy $E_f$ (eV) | Binding energy $E_b$ (eV) |
|---|---|---|
| V$_O$ chain | 6.80 | 0.39 |
| V$_O$ chain with Au | 5.31 | 1.88 |
| V$_O$ chain with Pt | 5.38 | 1.80 |
| V$_O$ chain with Pd | 5.66 | 1.54 |
| V$_O$ chain with Cu | 5.83 | 1.36 |
| V$_O$ chain with Ti | 5.96 | 1.24 |
| V$_O$ chain with Al | 5.78 | 1.41 |

**Supplementary Table 2. V$_O$ chain formation energies and binding energies (per V$_O$) in undoped and doped Al$_2$O$_3$**

However, it must be noted that the formation and binding energies for the various doped V$_O$ chain cases are not as drastically different as compared to the single V$_O$ NN to the dopant or that of the V$_O$ cluster. This is because the effect of the dopant is highly local. Since the chain has eight V$_O$ and just one dopant, the difference in energies, while evident, is not as significant. More details regarding this local effect of dopants on V$_O$ formation and the variation of chain formation and binding energies with the dopant-V$_O$ ratio is elaborated upon below.

**Local effect of dopant**

Supplementary Table 3 tabulates the formation energy of successively removing one $V_O$ (with increasing distance from the dopant) while forming the $V_O$ chain in doped $Al_2O_3$. It can be seen that in all the doped cases, the energy required to remove the second $V_O$ is significantly higher than the first and the energy successively increases as more $V_O$ are removed along the chain. However, the trend is preserved irrespective of the number of $V_O$ present, with Au, Pt and Pd cases always having lower $V_O$ formation energies as compared to Cu, Ti and Al.

| Number of $V_O$ | Au | Pt | Pd | Cu | Ti | Al |
|---|---|---|---|---|---|---|
| 1 | 0.64 | 0.47 | 1.91 | 3.62 | 4.38 | 3.63 |
| 2 | 3.47 | 3.37 | 4.13 | 5.01 | 5.06 | 5.04 |
| 3 | 3.64 | 3.56 | 4.22 | 4.95 | 5.15 | 5.08 |
| 4 | 4.32 | 4.2 | 4.73 | 5.29 | 5.36 | 5.40 |
| 5 | 4.75 | 4.69 | 5.12 | 5.54 | 5.55 | 5.53 |
| 6 | 5.07 | 4.99 | 5.35 | 5.69 | 5.74 | 5.66 |
| 7 | 5.20 | 5.25 | 5.55 | 5.84 | 5.86 | 5.77 |
| 8 | 5.31 | 5.39 | 5.66 | 5.83 | 5.96 | 5.79 |

**Supplementary Table 3. Successive oxygen vacancy ($V_O$) formation energies (per $V_O$) along the chain in interstitially doped $Al_2O_3$**

Supplementary Table 4 compares the $V_O$ chain formation and binding energy (with 8 $V_O$), where the chain has one vs. two dopant atoms at interstitial sites along the chain. For the 2-dopant atom case, chain formation energy for the Au, Pt and Pd cases is significantly lower than the Cu, Ti and Al cases (as compared to the 1 dopant case), again highlighting the local effect of the dopant. Similarly, the chain binding energy for the 2-dopant case is significantly much higher for Au, Pt and Pd cases.

| Dopant | Formation energy (eV) with 1 dopant/2 dopant atoms | Binding energy (eV) with 1 dopant/2 dopant atoms |
|---|---|---|
| None | 6.81 | 0.39 |
| Au | 5.31/4.10 | 1.88/3.45 |
| Pt | 5.39/3.81 | 1.80/3.21 |
| Pd | 5.66/4.42 | 1.54/2.67 |
| Cu | 5.83/5.032 | 1.36/2.39 |
| Ti | 5.96/5.16 | 1.24/2.02 |
| Al | 5.79/5.03 | 1.41/2.34 |

**Supplementary Table 4. Oxygen vacancy ($V_O$) chain formation energies and binding energies (per $V_O$) with one and two dopants at interstitial sites along the chain**

The initial structure of the Au-doped $V_O$ chain with Au in the interstitial site in $Al_2O_3$ is shown in Supplementary Figure 6b, with the positions of $V_O$ marked. The partial charge density of all the defect states within the band gap of the relaxed, undoped system, as shown in Supplementary Figure 6c reveals that a localized, conductive channel is formed along the vacancy chain direction. Such channel formation was also confirmed in a previous study of $V_O$ filament in $Al_2O_3$ ReRAMs[1], which found it to be comprised of Al 3s and 3p states. Analyzing the electronic density of states as shown in Supplementary Figure 6a, it can be seen that the oxygen vacancy chain reduces the band gap of $Al_2O_3$ by introducing overlapping mid-gap states. Introducing Au further reduces the band gap. This reduction in band gap implies that Al2O3 is locally metallic along the $V_O$ chain.

**Total density of states of $V_O$ chains with interstitial dopants at the center of the chain**

Supplementary Figure 7 shows the total density of states of $V_O$ chains with interstitial dopants at the center of the chain. The dopants serve to further reduce the band gap. No significant difference is seen between the various dopants, as the DOS profile is dominated primarily by the $V_O$, which are 8 in number.

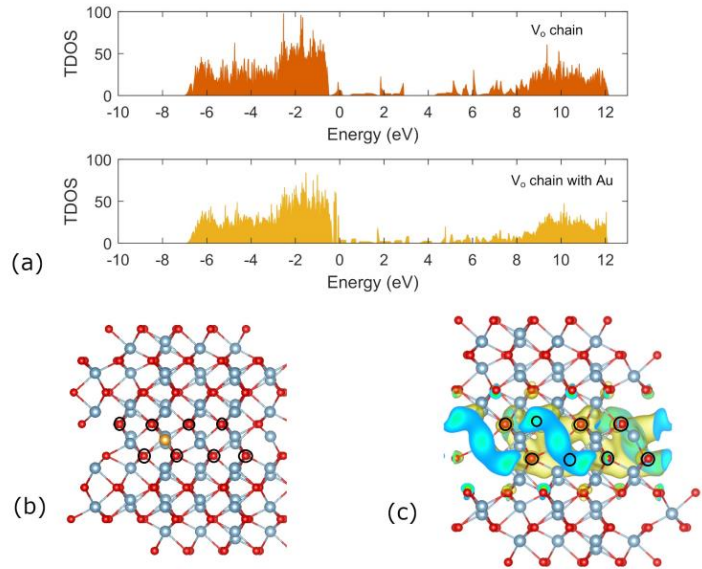

**Supplementary Figure 6. (a)** Total density of states (TDOS) of undoped system with $V_O$ chain and Au-doped system with $V_O$ chain **(b)** Initial structure of $V_O$ chain in Au-doped $Al_2O_3$ **(c)** Band decomposed charge density profile of states within the bandgap for the relaxed undoped $V_O$ chain system. Position of $V_O$ is marked by black circle

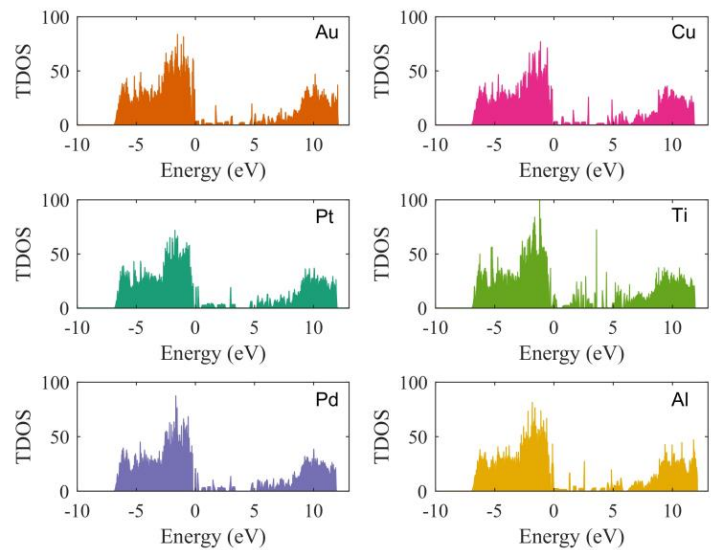

**Supplementary Figure 7.** Total density of states of $V_O$ chains in interstitially doped $Al_2O_3$

# VI. Migration energy studies in undoped and Au-doped $Al_2O_3$

We performed Climbing Image Nudged Elastic Band calculations[6,7] of neutral $V_O$ migration in undoped and Au-doped $Al_2O_3$. For the Au-doped case, we considered a system with a $V_O$ present at the nearest neighbor site of the dopant, and calculated the migration energies of an additional $V_O$ (at different distances from the [Au-$V_O$] complex). The migration paths considered are shown in Supplementary Figure 8, and the migration energies are tabulated in Supplementary Tables 5 and 6. The minimum energy pathways are shown in Supplementary Figs. 9 and 10. It is seen that the migration energy in the Au-doped system is ~0.5 eV lower that the corresponding jump in the undoped system.

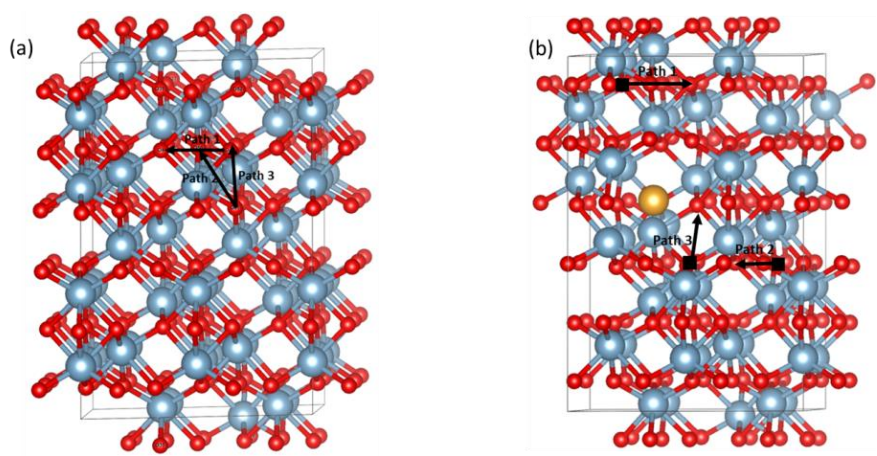

**Supplementary Figure 8.** $V_O$ migration pathways considered in (a) undoped $Al_2O_3$ and (b) Au-doped $Al_2O_3$

|  | Path 1 (2.55 Å) | Path 2 (2.64 Å) | Path 3 (2.75 Å) |
|---|---|---|---|
| $V_O$ migration energy | 3.59 eV (3.90 eV) | 4.67 eV (4.70 eV) | 4.10 eV (4.33 eV) |

**Supplementary Table 5.** $V_O$ migration energies in undoped $Al_2O_3$ (energy in parenthesis is from Ref. Yang, Moon Young, et al. *Applied Physics Letters* 103.9 (2013): 093504.)[1]

|  | Initial distance of $V_O$ from Au | Final distance of $V_O$ from Au | Jump distance | Migration energy |
|---|---|---|---|---|
| Path 1 | 4.99 Å | 4.94 Å | 2.54 Å | 3.09 eV |
| Path 2 | 4.93 Å | 4.81 Å | 2.54 Å | 3.21 eV |

| Path 3 | 2.81 Å | 2.69 Å | 2.64 Å | 4.13 eV |

**Supplementary Table 6.** $V_O$ migration energies in Au-doped $Al_2O_3$

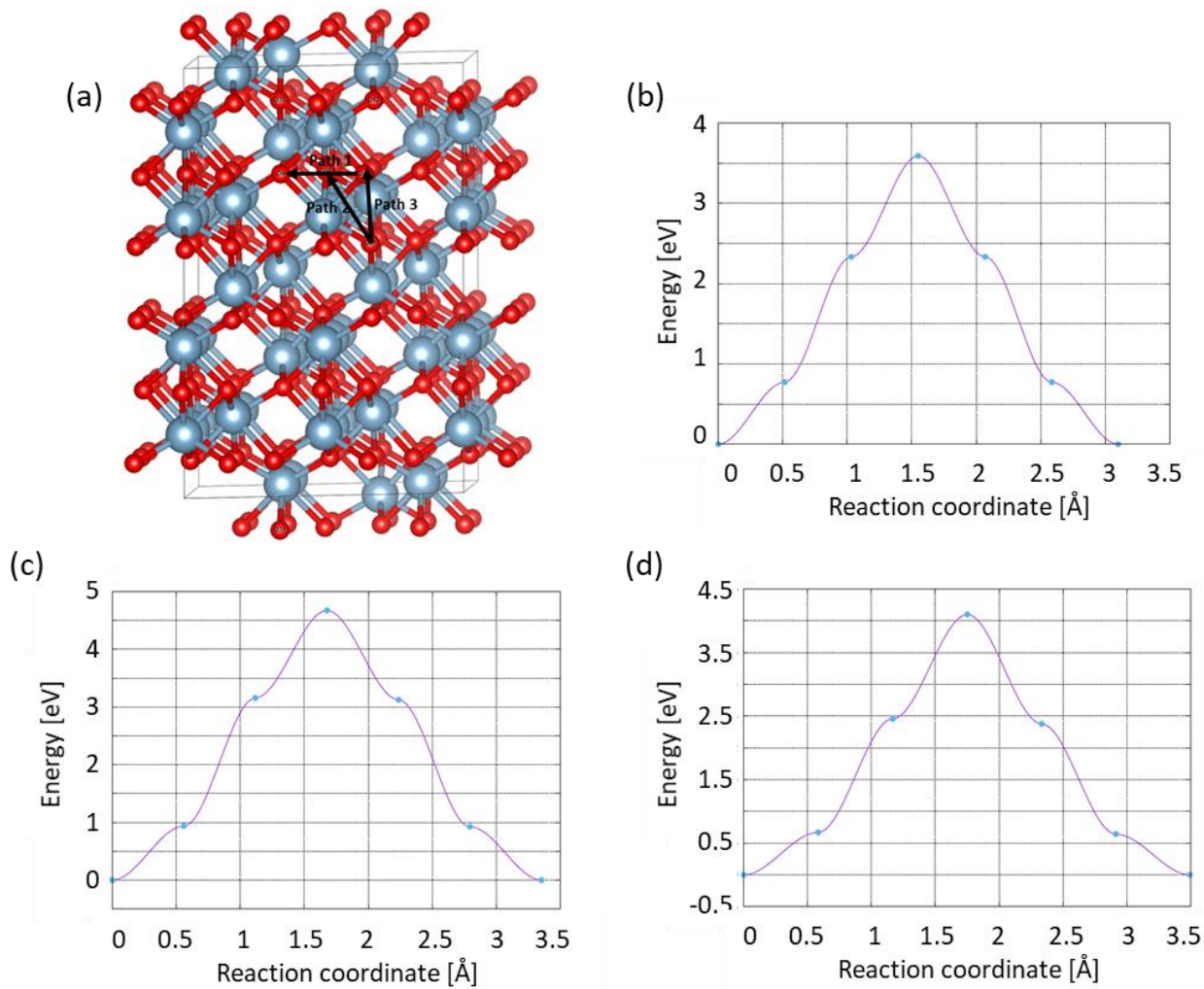

**Supplementary Figure 9.** Minimum energy pathways for $V_O$ migration in undoped $Al_2O_3$. Fitted spline curves are produced with VTST tools[6,7].

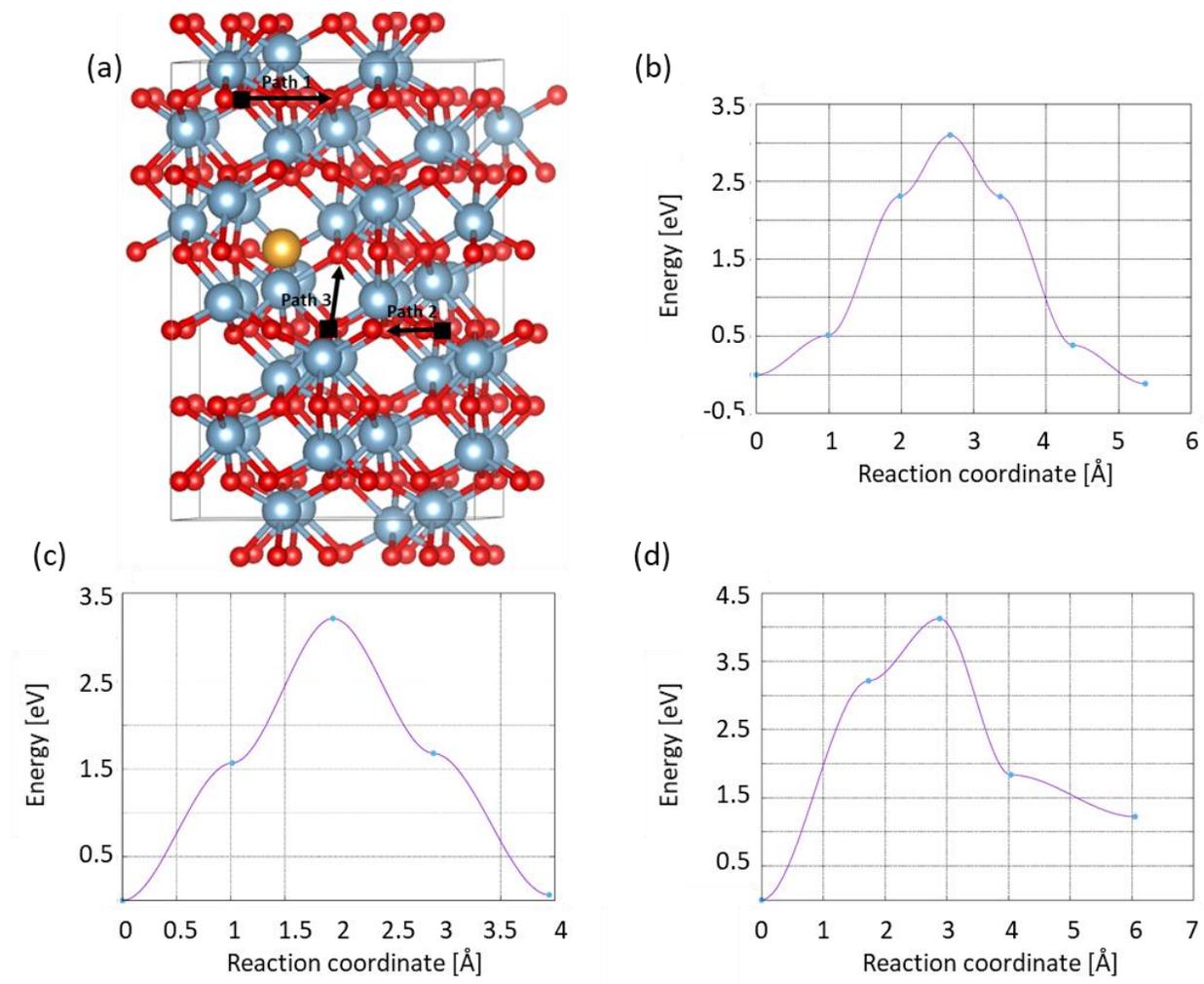

**Supplementary Figure 10.** Minimum energy pathways for $V_O$ migration in Au-doped $Al_2O_3$. Fitted spline curves are produced with VTST tools[6,7].

## VII. Bader charge analysis

Supplementary Table 7 table lists the Bader charges[8,9] on the dopant, nearest neighbor Al and nearest neighbor O. The Bader charge listed is the 'effective charge', i.e., the amount of charge transferred to (negative) or from (positive) the atom. Note that in undoped $Al_2O_3$ (or on Al and O farther away from the dopant), the Bader charge on Al and O is 2.46 e and -1.65 e respectively. It is seen that the noble metals Au, Pt and Pd all gain some amount of charge, and the nearest neighbor O does not gain as much charge as in undoped $Al_2O_3$ (or as compared to O farther away from the dopant). This weakens nearest neighbor Al-O bonds, facilitating easy $V_O$ formation, as discussed in the main text.

| Dopant | Bader charge on dopant | Bader charge on NN Al | Bader charge on NN O |
|---|---|---|---|
| Au | -0.41 | 2.42 | -1.56 |
| Pt | -0.59 | 2.46 | -1.54 |
| Pd | -0.36 | 2.44 | -1.57 |
| Cu | 0.39 | 2.43 | -1.61 |
| Ti | 0.54 | 2.28 | -1.65 |
| Al | 1.87 | 1.69 | -1.67 |

**Supplementary Table 7. Bader charge (*e*) on the dopant, nearest neighbor Al and nearest neighbor O**

## VIII. Local density of states of all interstitial dopants

Note how Au, Pt and Pd introduce states near the valence band maximum (VBM, which is at 0 eV) and multiple low-lying mid-gap states, unlike Cu, Ti and Al.

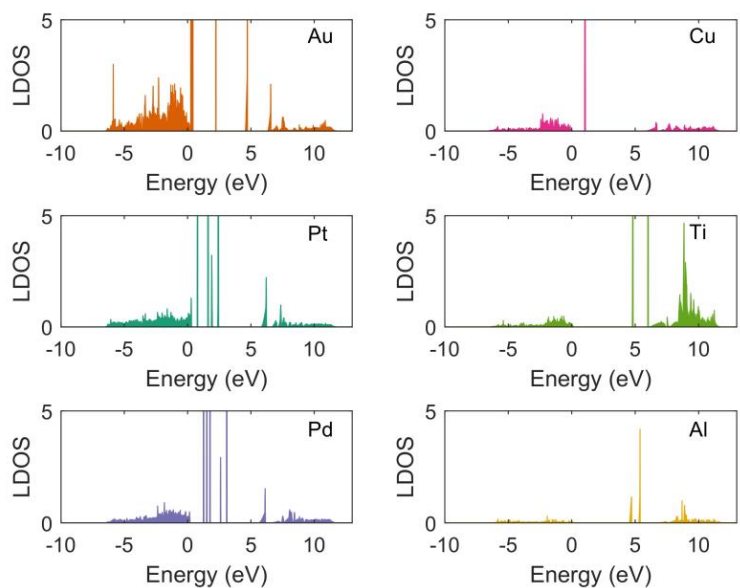

**Supplementary Figure 11. Local density of states on interstitial dopant in doped $Al_2O_3$. Valence band maximum (VBM) is at 0 eV.**

# IX. Device illustration

Illustrations of the 3-layer oxide Au-implanted resistive switching device is shown in Supplementary Figure at various length scales for clarity.

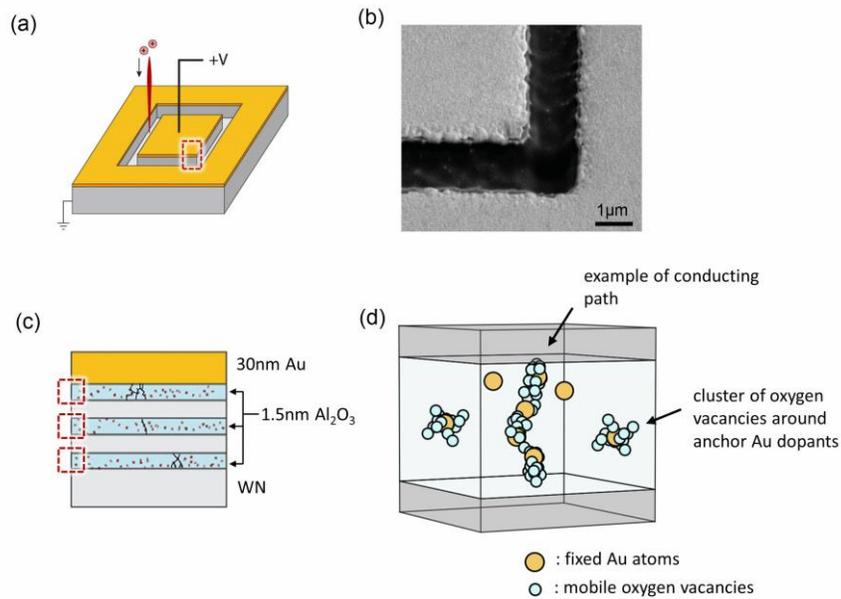

**Supplementary Figure 12. (a) 3D perspective view of the FIB milling carried out to define the area of a single resistive switching device.**
**(b) SEM image of region outlined with red dashed lines in (a). The darker region shows the exposed Si substrate while the lighter regions are the top Au film.**
**(c) Stack schematic showing the layers deposited on a Si substrate. The FIB milling implants Au atoms sideways into the $Al_2O_3$ layers. An atomistic schematic of the regions outlined in red dashed lines is shown in (d).**
**(d) The Au atoms act as anchors in gathering oxygen vacancies. Such clusters can potentially interact to form filaments or webs of filaments so that a conductive path can form across the entire oxide layer.**

# X. TEM imaging of ALD deposited thin films

TEM imaging was carried out on a typical 3 oxide layer device after performing TEM sample preparation with to extract a lamella for imaging (see Supplementary Figure ).

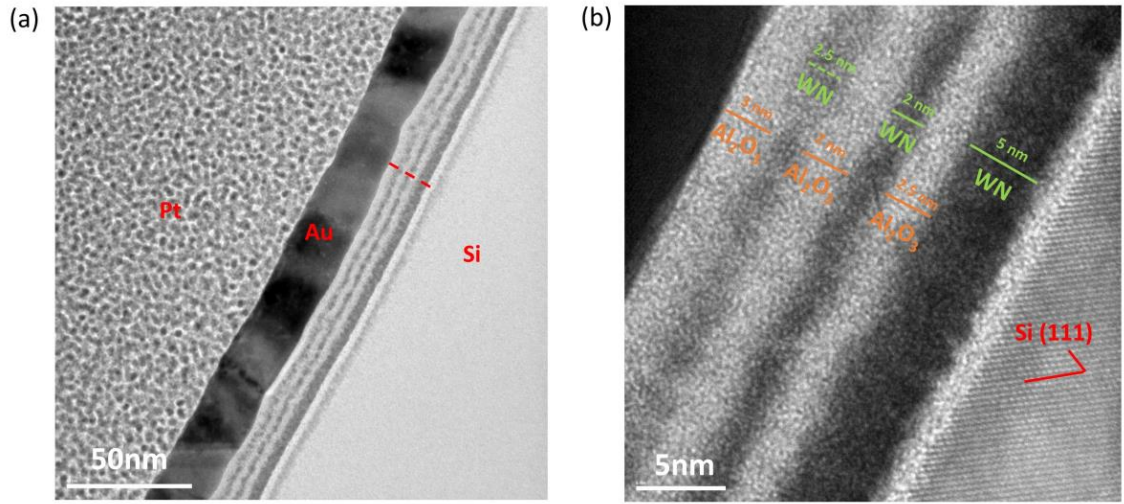

**Supplementary Figure 13. (a) Overview of imaged section showing the ALD deposited WN and Al$_2$O$_3$ layer pairs followed by a sputtered 30nm Au layer. (b) The thicker first layer of WN acts as the bottom electrode since it is conductive. Despite exposing the Si substrate to HF just before the ALD of Al$_2$O$_3$ and WN, a 1.5nm thick layer of native SiO$_2$ still grew. The interface between WN and Al$_2$O$_3$ appears to become more fuzzy with increasing repetition number, but each of the 3 deposited Al$_2$O$_3$ appears to be continuous while being separated from one another by intercalating WN layers.**

# XI. Aging and annealing observation

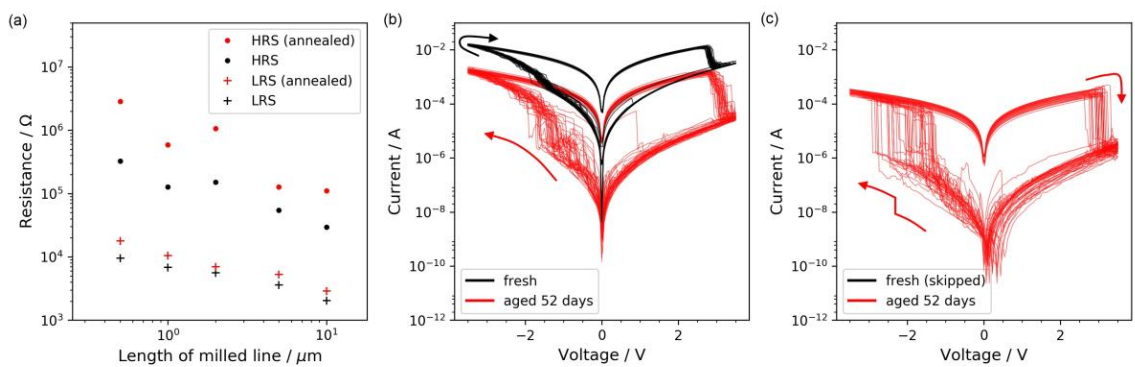

**Supplementary Figure 14. (a)** Devices annealed at 300°C for 3h become more resistive both in the HRS and LRS. These devices were FIB-processed by milling a line of variable length to activate the devices for switching. **(b)** A device was characterised after FIB-processing and also after room temperature aging. The resistances of both HRS and LRS as expected, but the gradual SET character at V<0 is retained. **(c)** When a device is aged without first conducting I-V sweeps on it, it ages differently and now loses its gradual SET character. The differences in aging in (b) and (c) show that multiple filamentary constituents are likely to be present.

The concurrent presence of two filamentary constituents in our devices, namely the Au dopant atoms and the oxygen vacancies, can be inferred from aging results as well. After a device was annealed at 300°C for 3 hours, the resistance of HRS increased by about 8x (see Supplementary Figure 4a). This is expected from previous studies and is thought to be due to filamentary material migrating out of the oxide layer[10]. More interesting, the effects of a two months room-temperature aging differs depending on whether the device was left to age after conducting I-V measurements on it, or before any measurement (i.e, untouched since the FIB-processing). When the device was aged between I-V measurements, the aged device shows the same character of a gradual SET and abrupt RESET albeit with both HRS and LRS becoming more resistive as expected (see Supplementary Figure 4b). However, if the device was aged after FIB-processing without first an initial I-V measurement, the aged device then shows abrupt SET and RESET transitions uncharacteristic of our typical device (see Supplementary Figure 4c). This can be understood in the context of the DFT studies described above where the Au dopant atoms can act as reservoirs for oxygen vacancies. When the device are switched before letting to age, the Au dopants atoms are stabilised and pinned in place by the surrounding oxygen vacancies, and so will retain a similar switching character with time. On the other hand, a device left to age straight after the FIB-processing will see the implanted Au atoms phase segregating out of the oxide and the subsequent switching character loses the identifiable gradual SET character distinct in our Au-implanted device.

# XII. Electroforming not required

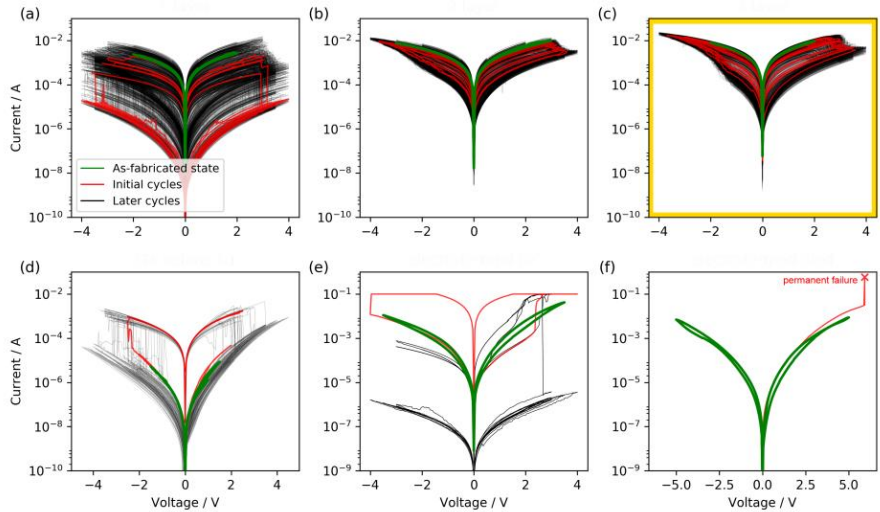

**Supplementary Figure 15. (a)** FIB-processed device with 1-layer 6nm oxide with electroforming not required but with subsequent inconsistent switching. **(b)** FIB-processed device with 2-layer oxide with 3.0nm per layer. **(c)** FIB-processed device with 3-layer oxide with 2.0nm per layer with no electroforming required. **(d)** Device with FIB mill processing done before Au layer deposition. **(e)** Device with e-beam deposited contact pads with a successful electroforming, but with following erratic switching cycles. **(f)** Device with e-beam deposited contact pads where the device blows up after electroforming and did not yield a functional device.

# XIII. Onset voltage distribution

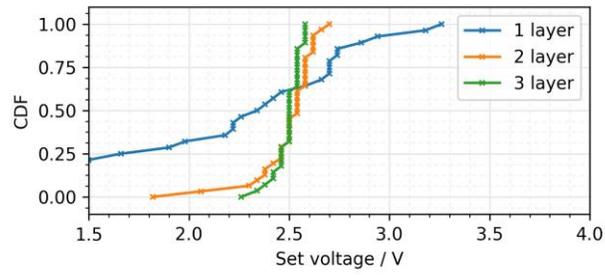

**Supplementary Figure 16.** The onset of RESET for 1,2,3-layers FIB-processed devices is reproduced as a cumulative distribution function shown here. Going from 1 to 2 to 3 oxide layers, the spread in the onset RESET voltage improves as it becomes narrower.

# XIV. Multibit switching scheme

Our device has good cycle-to-cycle consistency and a gradual SET, which makes the device good for demonstrating multibit switching as Supplementary Figure 7 shows. Another possible method to demonstrate multibit switching is by the use of SET or RESET pulses to induce incremental resistance changes (see Supplementary Figure 8).

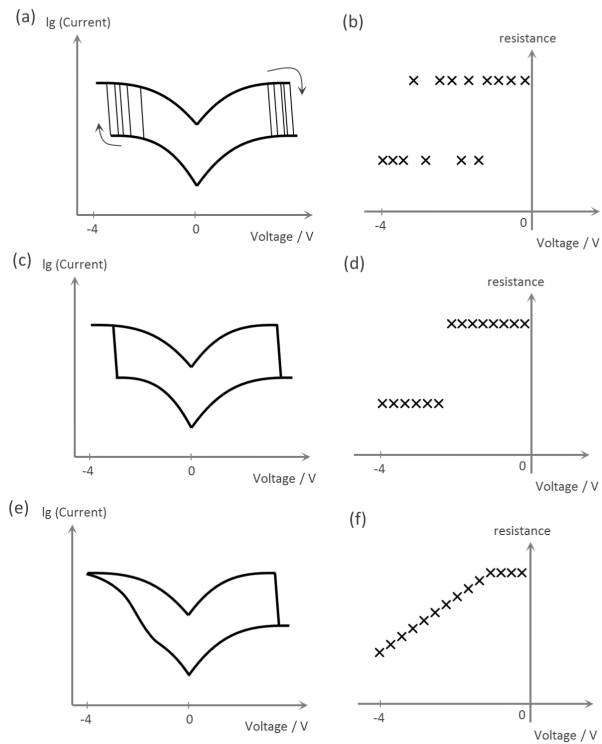

**Supplementary Figure 17. For a device with an abrupt SET and an inconsistent cycle-to-cycle SET voltage as shown in (a), it is not possible to use some particular SET stop voltage to put the device in a deterministic resistance state as shown by (b). For a device with an abrupt SET and a consistent cycle-to-cycle SET voltage as shown in (c), it is possible to use some particular SET stop voltage to put the device in one of 2 possible states shown by (d). For a device with a gradual SET and a consistent cycle-to-cycle SET voltage as shown in (e), it is possible use some particular SET stop voltage to put the device in one of many intermediate resistance states shown by (f).**

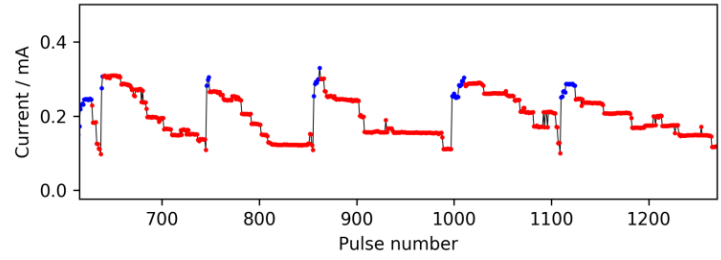

**Supplementary Figure 18. Response of device to 4V 50ns RESET pulses showing that incremental resistance changes are possible.**

# XV. Relaxed structures of $V_O$ at nearest neighbor site of interstitial dopant

Note how for Au, Pt and Pd cases, the dopant occupies the $V_O$ sites quite prominently.

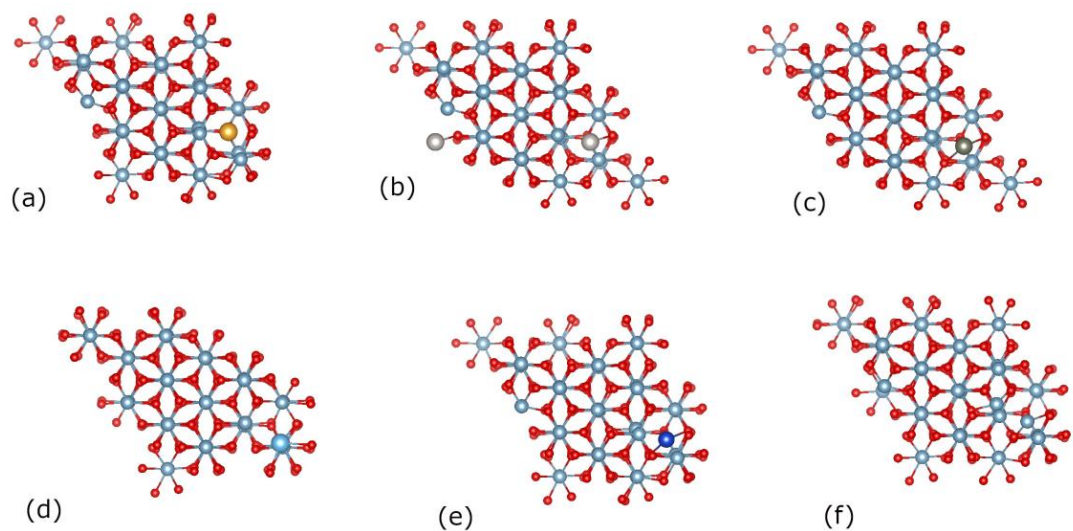

**Supplementary Figure 19. Relaxed structures of $V_O$ at nearest neighbor site of interstitial dopant (a) Au (b) Pt (c) Pd (d) Ti (e) Cu (f) Al.**

## XVI. Total density of states of doped $V_O$ clusters

Supplementary Figure 20 shows the total density of states of $V_O$ clusters formed by removing 4 $V_O$ at nearest neighbor sites of the interstitial dopant. Once again, the dopants serve to further reduce the band gap. No significant difference is seen between the various dopants, as the DOS profile is dominated primarily by the $V_O$, which are 4 in number.

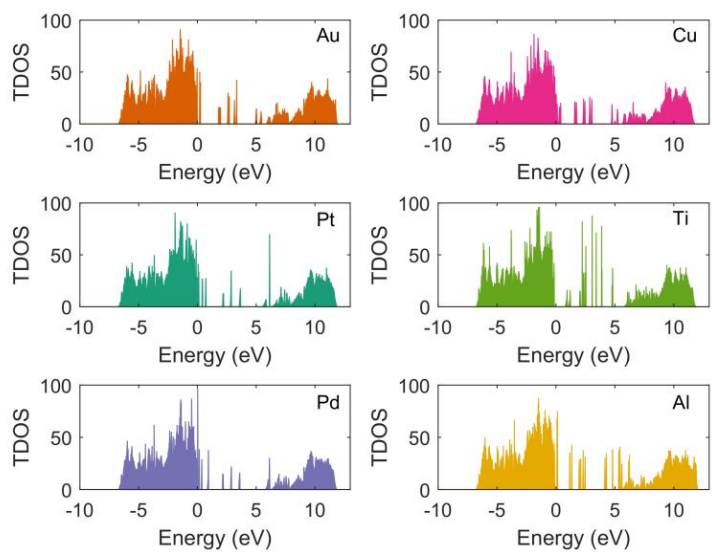

**Supplementary Figure 20. Total density of states of $V_O$ clusters in interstitially doped $Al_2O_3$**

## XVII. Simulated network model

Multilayer oxide films potentially contribute to a moderating effect on the voltages for SET and RESET through a network effect. This behavior can be obtained through an electrical network model where potential filament sites are treated as individual resistive elements that can be toggled between a high and low resistance state (HRS/LRS) (see Supplementary Figure). In the case that we have a physical web of interconnected filaments in each oxide layer, we can still approximate the web as a layer of filament sites in parallel. We find from simulations that this arrangement leads to an I-V behavior with a gradual SET and an abrupt RESET which is characteristic of our device.

This bare model is sufficient to show that a device with 2 or more oxide layer can have an abrupt RESET and a gradual SET (see Supplementary Figure 22a,b) despite each filament site having a threshold voltage with a large spread.

This chain effect at RESET and moderating effect at SET shows how a change from a single to multilayer device can reduce the dependence of the device's behavior on variable fabrication conditions and instead depend on the controllable device stack-up to give predictable resistive switching cycles. The moderating effect of the presence of multiple layers of filaments might be more pronounced if webs of filaments are present, as opposed to single filaments.

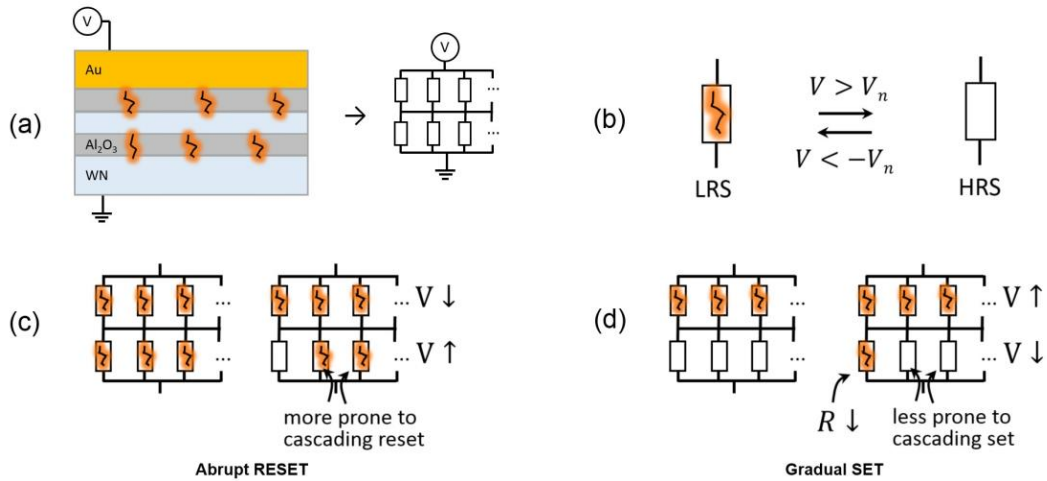

**Supplementary Figure 21. Schematic for network-level simulation of a multi-layer oxide structure.**
(a) A 2-layer oxide device was simulated. Each oxide layer was simulated to host $N$ number of filament sites, where each filament site can either host 1 or 0 filament corresponding to either the LRS or HRS. This network of filaments and their respective resistances can be converted into a network of resistors.
(b) Each filament site can be switched between the LRS or HRS based on whether the voltage across that filament site is above or below some preset threshold. The simulation randomizes each filament site $n \in [1, 2, 3, ..., N]$ to have a fixed threshold $V_n \in [V_1, V_2, V_3, ..., V_N]$ drawn from a random distribution.
(c) The top layer is assumed to have all filament sites fixed in the LRS. When the bottom layer starts to RESET, the total resistance of the bottom layer increases. This causes the voltage across the bottom layer to increase and creates a cascade of RESET actions. This creates an abrupt RESET transition.
(d) When the bottom layer starts to SET, the combined resistance of the bottom layer decreases and the voltage across the bottom layer drops. The applied voltage must be strengthened to compensate for this voltage drop before more filament sites can be SET. This creates a gradual SET transition.

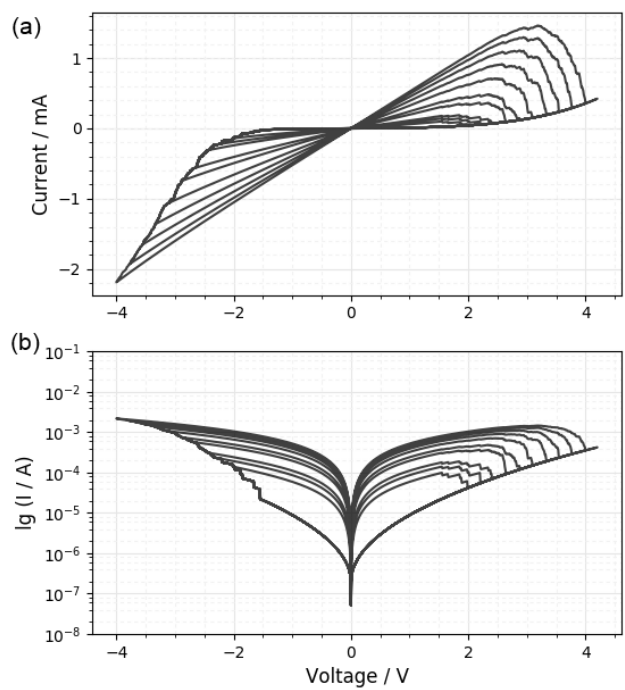

**Supplementary Figure 22.** (a) and (b) shows the linear and log plots of the voltage sweeps from the simulation. The simulation shows a gradual SET and a relatively abrupt RESET similar to what is observed experimentally.

## XVIII. Considerations for determination of implantation profile

The concentration of Au atoms in the $Al_2O_3$ is of high interest, but there lies many challenges to its determination regardless either via direct experimental observations or simulations. It is expected that Au dopants implanted sideways into $Al_2O_3$ as a side-effect of FIB will likely end up at shallow depths near the milled perimeter. However, the local concentration profile is difficult to pin down.

Experimental measurements of this dopant concentration is a challenge. X-ray photoelectron spectroscopy (XPS) has an atomic percentage detection limit of 0.1 to 1% for dopants which is likely satisfied by the *maximum* local concentration of the Au dopants. However, XPS has a millimeter beam size, which is several order of magnitudes larger than the sub-micrometer lengthscale over which the Au dopants will be implanted. Furthermore, the presence of a continuous Au film (as the top electrode) contributes a large background signal of Au, obliterating any chance to observe the signal from Au dopants within $Al_2O_3$. HRTEM is another candidate technique, but it will be at or beyond the limits of HRTEM to be able to identify sparse Au dopants amongst a background of amorphous $Al_2O_3$. Atomic probe tomography (APT) could be viable but currently not within our reach.

Simulations can be performed to estimate the dopant profile. Stopping and Range of Ions in Matter (SRIM) is a ubiquitous tool in nuclear engineering to predict the sputtering and implantation profile of energetic ionic bombardments into some material via Monte Carlo simulations. SRIM have been refined for several decades and corroborated by a vast amount of experimental data, so the simulation results themselves will be reliable. However, the limitations with SRIM is that only planar films are allowed, and dynamic simulations with geometries that is sputtered away with time is not possible. Several modeling assumptions is thus required to bridge SRIM simulations and the estimation of implantation profile in Supplementary Section I.

An alternative approach to more accurately estimate the concentration of Au ions will be to use directly implanted Au atoms with an Au ion source[11].This will reduce the number of modeling assumptions as compared to indirect Au implantation using a Ga ion source on Au film.

# Supplementary references